%% file: 0_main.tex
\documentclass{article}
\PassOptionsToPackage{numbers, compress}{natbib}
\usepackage[final]{neurips_2019}

\usepackage[utf8]{inputenc} % allow utf-8 input
\usepackage[T1]{fontenc}    % use 8-bit T1 fonts
\usepackage{hyperref}       % hyperlinks
\usepackage{url}            % simple URL typesetting
\usepackage{booktabs}       % professional-quality tables
\usepackage{amsfonts}       % blackboard math symbols
\usepackage{nicefrac}       % compact symbols for 1/2, etc.
\usepackage{microtype}      % microtypography
\usepackage{multicol}
\usepackage{color}
\usepackage{tikz}
\usetikzlibrary{bayesnet}

\usepackage{microtype}
\usepackage{graphicx}
\usepackage{caption}
\usepackage{subcaption}
\usepackage{booktabs} % for professional tables

\usepackage{mathtools}
\usepackage{amssymb}
\newcommand{\var}[1]{{\operatorname{#1}}}

\DeclareMathOperator*{\argmin}{\arg\!\min}
\newcommand{\mytilde}{\sim}
\newcommand{\KL}[2]{\mbox{KL}\left[ #1 \,\|\, #2 \right]}
\makeatletter
\newcases{mycases}{\quad}{%
  \hfil$\m@th\displaystyle{##}$}{$\m@th\displaystyle{##}$\hfil}{\lbrace}{.}
\makeatother

\usepackage[utf8]{inputenc}
\usepackage[english]{babel}

\usepackage{xcolor}
\usepackage{listings}

\title{Scalable Spike Source Localization in Extracellular Recordings using Amortized Variational Inference}

\author{%
Cole L. Hurwitz \\
School of Informatics\\
University of Edinburgh, United Kingdom\\
\texttt{cole.hurwitz@ed.ac.uk} \\
\And
Kai Xu \\
School of Informatics \\
University of Edinburgh, United Kingdom \\
\texttt{kai.xu@ed.ac.uk} \\
\AND 
Akash Srivastava \\
MIT–IBM Watson AI Lab \\
Cambridge, United States \\
\texttt{Akash.Srivastava@ibm.com} \\
\And
Alessio P. Buccino \\
Department of Informatics \\
University of Oslo, Oslo, Norway \\
\texttt{alessiob@ifi.uio.no} \\
\And
Matthias H. Hennig \\
School of Informatics \\
University of Edinburgh, United Kingdom \\
\texttt{m.hennig@ed.ac.uk} \\
}

\begin{document}

\maketitle

\begin{abstract}
Determining the positions of neurons in an extracellular recording is useful for investigating functional properties of the underlying neural circuitry. In this work, we present a Bayesian modelling approach for localizing the source of individual spikes on high-density, microelectrode arrays. To allow for scalable inference, we implement our model as a variational autoencoder and perform amortized variational inference. We evaluate our method on both biophysically realistic simulated and real extracellular datasets, demonstrating that it is more accurate than and can improve spike sorting performance over heuristic localization methods such as center of mass.

\end{abstract}

\input{includes/1_introduction}

\input{includes/2_spike_localization}

\input{includes/3_related_work}

\input{includes/4_method}

\input{includes/5_data_augmentation}
\input{includes/6_inference}

\input{includes/7_experiments}

\input{includes/8_results}

\input{includes/9_discussion.tex}

\newpage

% \bibliographystyle{rusnat}
% \bibliography{bibliography}

\newpage
\clearpage
\markright{Appendix}
\section*{Appendix}

\appendix
\input{includes/appendix}

\end{document}

%% file: includes/1_introduction.tex
\section{Introduction} \label{sec:intro}

Extracellular recordings, which measure local potential changes due to ionic currents flowing through cell membranes, are an essential source of data in experimental and clinical neuroscience. The most prominent signals in these recordings originate from action potentials (spikes), the all or none events neurons produce in response to inputs and transmit as outputs to other neurons. Traditionally, a small number of electrodes (channels) are used to monitor spiking activity from a few neurons simultaneously. Recent progress in microfabrication now allows for extracellular recordings from thousands of neurons using microelectrode arrays (MEAs), which have thousands of closely spaced electrodes \citep{Eversmann2003,Berdondini2005a,Frey2010,Ballini2014,Muller2015,yuan2016microelectrode,lopez201622,jun2017fully,dimitriadis2018not}. These recordings provide insights that cannot be obtained by pooling multiple single-electrode recordings \cite{kelly2007comparison}. This is a significant development as it enables systematic investigations of large circuits of neurons to better understand their function and structure, as well as how they are affected by injury, disease, and pharmacological interventions \cite{hennig2018scaling}.

On dense MEAs, each recording channel may record spikes from multiple, nearby neurons, while each neuron may leave an extracellular footprint on multiple channels. Inferring the spiking activity of individual neurons, a task called \textit{spike sorting}, is therefore a challenging blind source separation problem, complicated by the large volume of recorded data \cite{Rey2015}. Despite the challenges presented by spike sorting large-scale recordings, its importance cannot be overstated as it has been shown that isolating the activity of individual neurons is essential to understanding brain function \cite{miller2008all}. Recent efforts have concentrated on providing scalable spike sorting algorithms for large scale MEAs and already several methods can be used for recordings taken from hundreds to thousands of channels  \cite{pachitariu2016fast,lee2017yass,chung2017fully,yger2018spike,hilgen2017unsupervised,jun2017real}. However, scalability, and in particular automation, of spike sorting pipelines remains challenging \cite{carlson2019continuing}. 

One strategy for spike sorting on dense MEAs is to spatially localize detected spikes before clustering. In theory, spikes from the same neuron should be localized to the same region of the recording area (near the cell body of the firing neuron), providing discriminatory, low-dimensional features for each spike that can be utilized with efficient density-based clustering algorithms to sort large data sets with tens of millions of detected spikes \cite{hilgen2017unsupervised,jun2017real}. These location estimates, while useful for spike sorting, can also be exploited in downstream analyses, for instance to register recorded neurons with anatomical information or to identify the same units from trial to trial \cite{chelaru2005spike, hilgen2017unsupervised, Obien2015}. 

Despite the potential benefits of localization, preexisting methods have a number of limitations. First, most methods are designed for low-channel count recording devices, making them difficult to use with dense MEAs \cite{chelaru2005spike, somogyvari2005model, blanche2005polytrodes, lee2007efficient, kubo20083d, mechler2011three, mechler2012dipole, somogyvari2012localization}. Second, current methods for dense MEAs utilize cleaned extracellular action potentials (through spike-triggered averaging), disallowing their use before spike sorting \cite{ruz2014localising, buccino2018combining}. Third, all current model-based methods, to our knowledge, are non-Bayesian, relying primarily on numerical optimization methods to infer the underlying parameters. Given these current limitations, the only localization methods used consistently before spike sorting are simple heuristics such as a center of mass calculation \cite{nadasdy1998extracellular, Prentice2011, hilgen2017unsupervised, jun2017real}.

% Due to these reasons, the only localization methods used consistently before spike sorting are simple heuristics such as a center of mass calculation \cite{nadasdy1998extracellular, Prentice2011, hilgen2017unsupervised, jun2017real}.

% Existing \textit{pre-sorting} localization methods use simple heuristics such as a center of mass calculation. There have been much work on estimating cell body locations \textit{post-sorting}, but these methods either use supervised learning and are thus unsuitable when the origin of the spikes is unknown or are too computationally expensive to apply to large-scale datasets \cite{ruz2014localising, mechler2011three, buccino2018combining}.Post-sorting localization methods can be biased by spike sorting errors, making them less reliable in real experimental settings \cite{buccino2018combining}.
In this paper, we present a scalable Bayesian modelling approach for spike localization on dense MEAs (less than $\sim50\mu$m between channels) that can be performed \textit{before} spike sorting. Our method consists of a generative model, a data augmentation scheme, and an amortized variational inference method implemented with a variational autoencoder (VAE) \cite{dayan1995helmholtz, kingma2013auto, rezende2014stochastic}. Amortized variational inference has been used in neuroscience for applications such as predicting action potentials from calcium imaging data \cite{speiser2017fast} and recovering latent dynamics from single-trial neural spiking data \cite{pandarinath2018inferring}, however, to our knowledge, it has not been used in applications to extracellular recordings.

After training, our method allows for localization of one million spikes (from high-density MEAs) in approximately 37 seconds on a TITAN X GPU, enabling real-time analysis of massive extracellular datasets. To evaluate our method, we use biophysically realistic simulated data, demonstrating that our localization performance is significantly better than the center of mass baseline and can lead to higher-accuracy spike sorting results across multiple probe geometries and noise levels. We also show that our trained VAE can generalize to recordings on which it was not trained. To demonstrate the applicability of our method to real data, we assess our method qualitatively on real extracellular datasets from a Neuropixels \cite{jun2017fully} probe and from a BioCam4096 recording platform. 

To clarify, our contribution is not full spike sorting solution. Although we envision that our method can be used to improve spike sorting algorithms that currently rely center of mass location estimates, interfacing with and evaluating these algorithms was beyond the scope of our paper.

% We quantitatively evaluate our method on biophysically realistic simulated data demonstrating that our localization performance is significantly better than the center of mass baseline and can lead to high-accuracy spike sorting results across multiple probe geometries and noise levels. We also show that our trained VAE can generalize to recordings on which it was not trained. Finally, to demonstrate the applicability of our method to real data, we assess our method qualitatively on data samples from a Neuropixels probe and from a BioCam4096 recording platform.

%% file: includes/2_spike_localization.tex
\section{Background} \label{sec:background}

\subsection{Spike localization} \label{sec:spikeloc}

% In this section, we introduce some notation and definitions relevant for spike localization. We then describe the task of spike localization using our defined notation.
We start by introducing the relevant notation. Let $\boldsymbol{s} \coloneqq \{s_n\}_{n=1}^N$ be the set of $N$ spikes that are detected in an extracellular recording. For each spike, we define the source location of the spike to be $p_{s_n} \coloneqq (x_{s_n} y_{s_n}, z_{s_n}) \in \mathbb{R}^3$. We further denote $p_{\boldsymbol{c}} \coloneqq \{p_{c_m}\}_{m=1}^M$ to be the position of all $M$ channels on the MEA. All positions are relative to the origin, which we set to be the center of the MEA. When a spike occurs during an extracellular recording, we assume there is a stereotypical spatiotemporal pattern that is recorded on the channels in the array, i.e., the recorded extracellular waveforms. The recorded extracellular waveforms of a spike $s_{n}$ on a channel $c_m$ can then be defined as $w_{n,m} \coloneqq \{r_{n,m}^{(0)}, r_{n,m}^{(1)}, ..., r_{n,m}^{(t)}, ..., r_{n,m}^{(T)}\}$ where $r_{n,m}^{(t)} \in \mathbb{R}$ and $t = 0,\dots,T$.

Localizing a spike can now be defined as follows: \textit{Localizing a spike $s_{n}$ is equivalent to solving for the corresponding point source location $p_{s_{n}}$ given the observed waveforms $\boldsymbol{w}_{n}$ and the channel positions $p_{\boldsymbol{c}}$.}

We perform source localization independently for each spike. Crucially, we make the assumption that the point source location $p_{s_n}$ of a spike is approximately the location of the firing neuron's soma. This is a useful assumption as it allows us to localize the underlying neurons' positions without spike sorting. We discuss limitations of this modelling assumption in the Discussion section.

% \footnote{We make the assumption that the true location of the point source location, $p_{s_{i,k}}$ is actually the location of the firing neuron's soma, $p_{n_{i}}$.
% Given the complex morphological structure of many neurons, this assumption may not be correct, but it provides a simple way to assess localization performance and improve future models.}

%% file: includes/3_related_work.tex
\subsection{Center of mass} \label{sec:com}

% \section{Related Work}

Many modern spike sorting algorithms localize spikes on MEAs using the center of mass or barycenter method \cite{Prentice2011,hilgen2017unsupervised,jun2017real}. We summarize the traditional steps for localizing a spike, $s_{n}$ using this method.
First, let us define 
$
\alpha_{n} \coloneqq \min_t w_{n,m}
$
as the negative amplitude peak of the waveform, $w_{n,m}$, generated by $s_{n}$ and recorded on channel, $c_{m}$. We consider the negative peak amplitude as a matter of convention since spikes are defined as inward currents. Then, let $\boldsymbol{\alpha}_{n} \coloneqq (\alpha_{n,m})_{m=1}^M$ be the vector of all amplitudes generated by $s_{n}$ and recorded by all $M$ channels on the MEA.

To find the center of mass of a spike, $s_{n}$, the first step is to determine the central channel for the calculation. This central channel is set to be the channel which records the minimum amplitude for the spike, $c_{m_{min}} \coloneqq c_{\argmin_m \alpha_{n,m}}$
The second and final step is to take the $L$ closest channels to $c_{m_{min}}$ and compute,
$
\hat{x}_{s_{n}} = \dfrac{\sum_{l=1}^{L+1} (x_{c_l})\vert \alpha_{n,l} \vert}{\sum_{l=1}^{L+1} \vert \alpha_{n,l} \vert}, 
\hat{y}_{s_{n}} = \dfrac{\sum_{l=1}^{L+1} (y_{c_l})\vert \alpha_{n,j} \vert}{\sum_{l=1}^{L+1} \vert \alpha_{n,l} \vert}
$
where all of the $L + 1$ channels' positions and recorded amplitudes contribute to the center of mass calculation.

The center of mass method is inexpensive to compute and has been shown to give informative location estimates for spikes in both real and synthetic data \cite{Prentice2011,Muthmann2015,hilgen2017unsupervised,jun2017real}. Center of mass, however, suffers from two main drawbacks: First, since the chosen channels will form a convex hull, the center of mass location estimates must lie \textit{inside} the channels' locations, negatively impacting location estimates for neurons outside of the MEA. Second, center of mass is biased towards the chosen central channel, potentially leading to artificial separation of location estimates for spikes from the same neuron \cite{Prentice2011}.

% This bias can be alleviated by increasing the number of channels used in the location estimate at the cost of lower accuracy on spikes detected near the edge of the array and a higher risk of overlapping with other spikes. The decreased accuracy on edge spikes when increasing the number of channels stems from the increased bias towards the center of the array \cite{Prentice2011}.

% This is especially an issue for dense laminar probes, like Neuropixels, which have a narrow, long shank so that they can be inserted into a living brain \cite{jun2017fully}.

%% file: includes/4_method.tex
\section{Method} \label{sec:method}
In this section, we introduce our scalable, model-based approach to spike localization. We describe the generative model, the data augmentation procedure, and the inference methods.

\begin{figure}
\begin{subfigure}{.27\textwidth}
  \centering
  \tikz{
% nodes
 \node[obs] (w) {$w$};%
 \node[latent,above=of w,yshift=-.5cm, xshift=-1.25cm] (a) {$a$}; %
 \node[latent,above=of w,yshift=-.5cm, xshift=-.4cm] (x) {$x$}; %
 \node[latent,above=of w,yshift=-.5cm, xshift=.4cm] (y) {$y$}; %
 \node[latent,above=of w,yshift=-.5cm, xshift=1.25cm] (z) {$z$}; %
%  \draw (0,1.69) circle(.36cm);
% plate
%  \plate [inner sep=.2cm,yshift=.1cm] {plate1} {(x)(y)(z_{i})(z_{r})} {$T$}; %
%  \tikzset{plate caption/.style={caption, node distance=0, inner sep=0pt, below left=5pt and -35pt of #1.south,text height=2.5em,text depth=-10.0em}}
 \plate [inner sep=.2cm,yshift=.1cm] {plate0}
 {(w)(a)(x)(y)(z)} {$N$}; %
% edges
 \edge {a} {w}  
 \edge {x} {w} 
 \edge {y} {w} 
 \edge {z} {w} 
 }
  \centering
  \captionsetup{justification=centering}
  \caption{Graphical model}
  \label{fig:tndm_prob}
\end{subfigure}
\begin{subfigure}{.73\textwidth}
%   \centering
  \includegraphics[scale=.18]{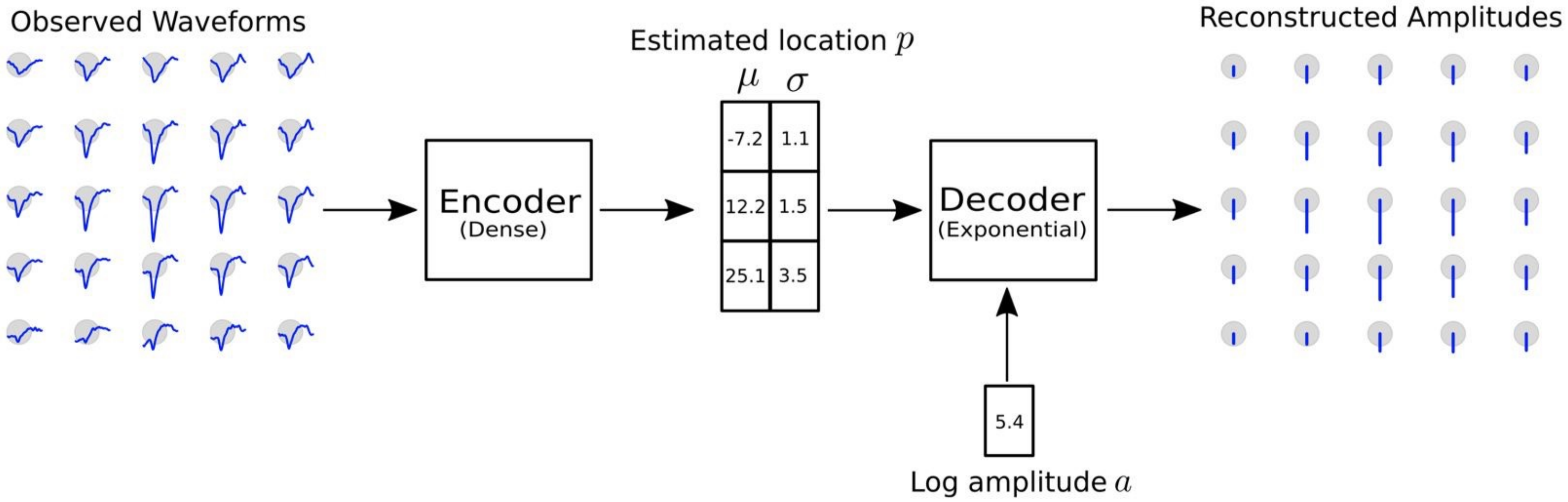}
  \centering
  \captionsetup{justification=centering}
  \caption{Architecture}
  \label{fig:dm_arch}
\end{subfigure}
\caption{(a) For each spike $s_n$, there are four latent variables that give rise to the observed waveforms $w_n$. These are the initial amplitude $a$ and the $x$, $y$, $z$ values of the source location $p_{s_n}$. Each spike is localized independently as indicated by the plate diagram. (b) Our model is implemented as a variational autoencoder with a dense encoder and an exponential function with a Gaussian observation model. The latent space consists of the three location variables $x_n$, $y_n$, $z_n$. The initial amplitude $a_n$ is inferred using a maximum likelihood estimate to improve stability.}
\label{fig:figure1}
\end{figure}

\subsection{Model} \label{sec:model}
Our model uses the recorded amplitudes on each channel to determine the most likely source location of $s_{n}$. We assume that the peak signal from a spike decays exponentially with the distance from the source, r: $a \exp(br)$ where $a, b \in \mathbb{R}, r \in \mathbb{R^+}$. This assumption is well-motivated by experimentally recorded extracellular potential decay in both a salamander and mouse retina \cite{segev2004recording,hilgen2017unsupervised}, as well as a cat cortex \cite{gray1995tetrodes}. It has also been further corroborated using realistic biophysical simulations \cite{Hagen2015}.

We utilize this exponential assumption to infer the source location of a spike, $s_{n}$, since localization is then equivalent to solving for  $s_{n}$'s unknown parameters, $\theta_{s_{n}} \coloneqq \{a_{n}, b_{n}, x_{s_{n}}, y_{s_{n}}, z_{s_{n}}\}$ given the observed amplitudes, $\boldsymbol{\alpha}_{n}$. To allow for localization without knowing the identity of the firing neuron, we assume that each spike has individual exponential decay parameters, $a_{n}, b_{n}$, and individual source locations, $p_{s_{n}}$. We find, however, that fixing $b_{n}$ for all spikes to a constant that is equal to an empirical estimate from literature (decay length of $\mytilde28 \mu m$) works best across multiple probe geometries and noise levels, so we did not infer the value for $b_{n}$ in our final method.
We will refer to the fixed decay rate as $b$ and exclude it from the unknown parameters moving forward.

The generative process 
of our exponential model is as follows,
\begin{equation}
\begin{aligned}
&a_{n} \mytilde N(\mu_{a_{n}}, \sigma_{a}),\;
x_{s_{n}} \mytilde N(\mu_{x_{s_{n}}}, \sigma_{x}),\;
y_{s_{n}} \mytilde N(\mu_{y_{s_{n}}}, \sigma_{y}),\;
z_{s_{n}} \mytilde N(\mu_{z_{s_{n}}}, \sigma_{z})\\
&\hat{\boldsymbol{r}}_{n} = \Vert (x_{s_{n}},y_{s_{n}},z_{s_{n}}) - p_{\boldsymbol{c}}\Vert_2,\;\boldsymbol{\alpha}_{n} \mytilde \mathcal{N} (a_{n}\exp(b\hat{\boldsymbol{r}}_{n}), I)
\end{aligned}
\end{equation}
In our observation model, the amplitudes are drawn from an isotropic Gaussian distribution with a variance of one. We chose this Gaussian observation model for computational simplicity and since it is convenient to work with when using VAEs. This is because learning the variance of the observation model in a VAE can be numerically unstable \cite{rybkin2021simple}.  We discuss the limitations of our modeling assumptions in Section~\ref{sec:discussion} and propose several extensions for future works. 

For our prior distributions, we were careful to set sensible parameter values. We found that inference, especially for a spike detected near the edge of the MEA, is sensitive to the mean of the prior distribution of $a_{n}$, therefore, we set $\mu_{a_{n}} = \lambda\alpha_{n,m_{min}}$ where $\alpha_{n,m_{min}}$ is the smallest negative amplitude peak of $s_{n}$. We choose this heuristic because the absolute value of $\alpha_{n,m_{min}}$ will always be smaller than the absolute value of the amplitude of the spike at the source location, due to potential decay. Therefore, scaling $\alpha_{n,m_{min}}$ by $\lambda$ gives a sensible value for $\mu_{a_{n}}$. We empirically choose $\lambda = 2$ for the final method after performing a grid search over $\lambda = \{1, 2, 3\}$. The parameter, $\sigma_{a}$, does not have a large affect on the inferred location so we set it to be approximately the standard deviation of the $\boldsymbol{\alpha}_{n,m_{min}}$ (50). The location prior means, $\mu_{x_{s_{n}}}, \mu_{y_{s_{n}}}, \mu_{z_{s_{n}}}$, are set to the location of the minimum amplitude channel, $p_{c_{m_{min}}}$, for the given spike. The location prior standard deviations, $\sigma_{x}, \sigma_{y}, \sigma_{z}$, are set to large constant values to flatten out the distributions since we do not want the location estimate to be overly biased towards $p_{c_{m_{min}}}$.

%% file: includes/5_data_augmentation.tex
\subsection{Data Augmentation} \label{sec:dataaug}
For localization to work well, the input channels should be centered around the peak spike, which is hard for spikes near the edges (edge spikes). To address this issue, we employ a two-step data augmentation. First, inputs for edge spikes are padded such that the channel with the largest amplitude is at the center of the inputs. Second, all channels are augmented with an indicating variable which provides signal to distinguish them for the inference network. To be more specific, we introduce virtual channels outside of the MEA which have the same layout as the real, recording channels (see appendix C). We refer to a virtual channel as an "unobserved" channel, $c_{m_{u}}$, and to a real channel on the MEA as an "observed" channel, $c_{m_{o}}$. We define the amplitude on an unobserved channel, $\alpha_{n,m_{u}}$, to be zero since unobserved channels do not actually record any signals. We let the amplitude for an observed channel, $\alpha_{n,m_{o}}$, be equal to $\min_t w_{n,m_{o}}$, as before.

Before defining the augmented dataset, we must first introduce an indicator function, $1_o: \alpha \rightarrow \{0, 1\}$:
\begin{align*}
1_o(\alpha) & = \begin{mycases}
                1, & \text{if $\alpha$ is from an observed channel},\\
                0, & \text{if $\alpha$ is from an unobserved channel}.
            \end{mycases}
\end{align*}
where $\alpha$ is an amplitude from any channel, observed or unobserved. 

To construct the augmented dataset for a spike, $s_{n}$, we take the set of $L$ channels that lie within a bounding box of width $W$ centered on the \textit{observed} channel with the minimum recorded amplitude, $c_{m_{o_{min}}}$. We define our newly augmented observed data for  $s_{n}$ as,
\begin{equation} \label{eq:11}
\boldsymbol{\beta}_{n} \coloneqq \{(\alpha_{n,l}, 1_o(\alpha_{n,l})\}_{l=1}^L
\end{equation}
So, for a single spike, we construct a $L\times2$ dimensional vector that contains amplitudes from $L$ channels and indices indicating whether the amplitudes came from observed or unobserved channels.

Since the prior location for each spike is at the center of the subset of channels used for the observed data, for edge spikes, the data augmentation \textit{puts the prior closer to the edge} and is, therefore, more informative for localizing spikes near/off the edge of the array. Also, since edge spikes are typically seen on less channels, the data augmentation serves to ignore channels which are away from the spike, which would otherwise be used 
if the augmentation is not employed.

%% file: includes/6_inference.tex
\subsection{Inference} \label{sec:inference}

Now that we have defined the generative process and data augmentation procedure, we would like to compute the posterior distribution for the unknown parameters of a spike, $s_{n}$,
\begin{alignat}{2}
p(a_{n}, x_{s_{n}}, y_{s_{n}}, z_{s_{n}} \vert \boldsymbol{\beta}_{n})
\end{alignat}
given the augmented dataset, $\boldsymbol{\beta}_{n}$. To infer the posterior distribution for each spike, we utilize two methods of Bayesian inference: MCMC sampling and amortized variational inference.

\begin{figure}[!t] 
\centering
\includegraphics[width=1\textwidth]{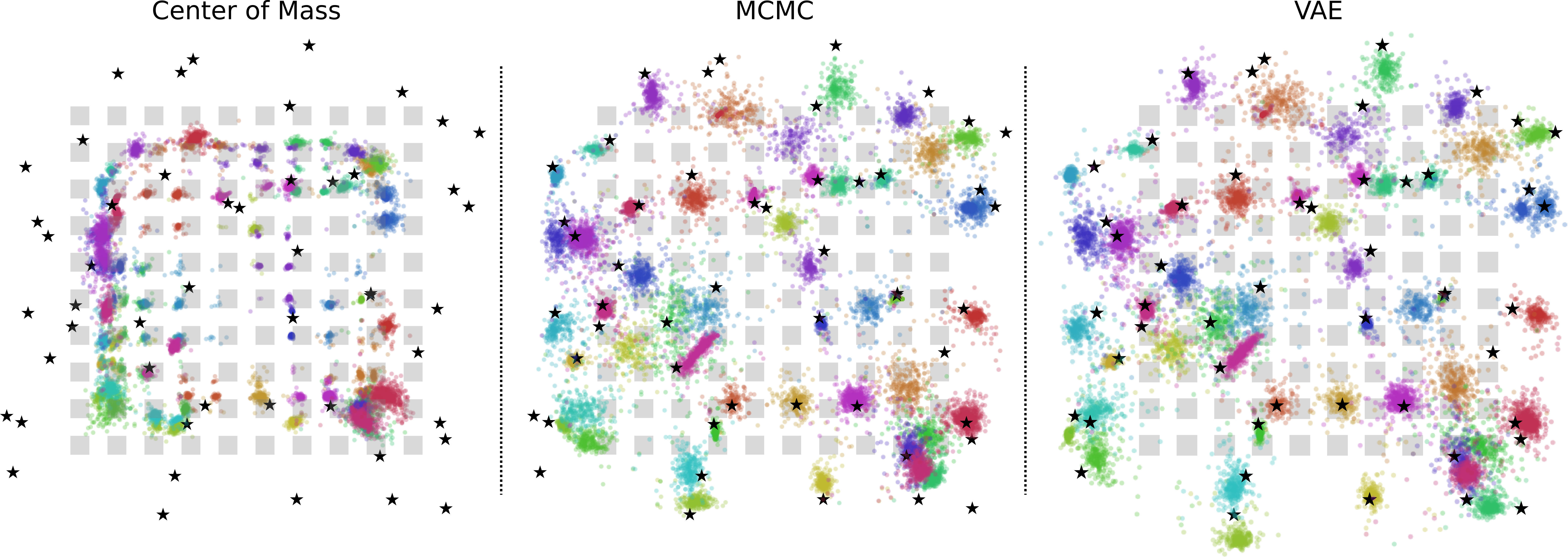}
\caption{\textit{Estimated spike locations for the different methods on a 10$\mu$V recording.} Center of mass estimates (left) are calculated using 16 observed amplitudes. The MCMC estimated locations (middle) used 9-25 observed amplitudes for inference, and the VAE model (right) was trained on 9-25 observed amplitudes and a 10 amplitude jitter (amplitude jitter is described in \ref{sec:ampjit}).}
\label{fig:simplot}
\vspace{-0.15in}
\end{figure}

\subsubsection{MCMC sampling} \label{sec:mcmc}
We use MCMC to assess the validity and applicability of our model to extracellular data. We implement our model in \textit{Turing} \cite{ge2018turing}, a probabilistic modeling language in Julia. We run Hamiltonian Monte Carlo (HMC) \cite{neal2011mcmc} for 10,000 iterations with a step size of 0.01 and a step number of 10. 
% To recover the posterior mean values for a spike, we take the average value of the associated chain.
We use the posterior means of the location distributions as the estimated location.\footnote{The code for our MCMC implementation is provided in Appendix H.}

% Interestingly, when we used the particle Gibbs sampler, it performed quite poorly on our model and does not properly explore the posterior. We assume this is because of the exponential function in the model causing numerical issues for particle Gibbs. 
Despite the ease use of probabilistic programming and asymptotically guaranteed inference quality of MCMC methods, the scalability of MCMC methods to large-scale datasets is limited. This leads us to implement our model as a VAE and to perform amortized variational inference for our final method.

\subsubsection{Amortized variational inference} \label{sec:AMI}
To speed up inference of the spike parameters, we construct a VAE and use amortized variational inference to estimate posterior distributions for each spike. In variational inference, instead of sampling from the target intractable posterior distribution of interest, we construct a variational distribution that is tractable and minimize the Kullback–Leibler (KL) divergence between the variational posterior and the true posterior. Minimizing the KL divergence is equivalent to maximizing the evidence lower bound (ELBO) for the log marginal likelihood of the data.
In VAEs, the parameters of the variational posterior are not optimized directly, but are, instead, computed by an inference network. When training the VAE, we found that inference of the initial amplitude $a_{n}$, especially for a spike detected near the edge of the MEA, is quite sensitive. To improve stability, we decided to do a maximum likelihood estimate for the mean of initial amplitude $\mu_{a_{n}}$ with a fixed variance $\sigma_{a}$. Therefore, we only define our variational posterior for the source location, $x_n,y_n,z_n$.

We define our variational posterior for $x_n,y_n,z_n$ as a multivariate Normal with diagonal covariance where the mean and diagonal of the covariance matrix are computed by an inference network
\begin{align}
q_\Phi(x_n,y_n,z_n) = \mathcal{N}(\pmb \mu_{\phi_1}(f_{\phi_0}(\upsilon_{n})), \pmb \sigma^2_{\phi_2}(f_{\phi_0}(\upsilon_{n})))
\end{align}
The inference network is implemented as a feed-forward, deep neural network parameterized by $\Phi = \{\phi_0,\phi_1,\phi_2\}$. As one can see, the variational parameters are a function of the input $\pmb \upsilon$.

When using an inference network, the input can be any part of the dataset so for our method, we use, $\boldsymbol{\upsilon_{n}}$, as the input for each spike, $s_{n}$, which is defined as follows:
\begin{equation} \label{eq:5}
\boldsymbol{\upsilon}_{n} \coloneqq \{(w_{n,l}, 1_o(\alpha_{n,l})\}_{l=1}^L
\end{equation}
where $w_{n,l}$ is the waveform detected on the lth channel (defined in Section \ref{sec:spikeloc}). Similar to our previous augmentation, the waveform for an unobserved channel is set to be all zeros. We choose to input the waveforms rather than the amplitudes because, empirically, it encourages the inferred location estimates for spikes from the same neuron to be better localized to the same region of the MEA. For both the real and simulated datasets, we used $\mytilde$2 ms of readings for each waveform.

The decoder for our method reconstructs the \textit{amplitudes} from the observed data rather than the waveforms. Since we assume an exponential decay for the amplitudes, the decoder is a simple Gaussian likelihood function, where given the Euclidean distance vector $\hat{\boldsymbol{r}_{n}}$, computed by samples from the variational posterior, the decoder reconstructs the mean value of the observed amplitudes with a fixed variance. The decoder is parameterized by the exponential parameters of the given spike, $s_{n}$, so it reconstructs the amplitudes of the augmented data, $\boldsymbol{\beta}_{n}^{(0)}$, with the following expression: $\hat{\boldsymbol{\beta}}_{n}^{(0)} \coloneqq a_{n}\exp(b\hat{\boldsymbol{r}}_{n})\times\beta_{n}^{1} \nonumber$ where $\hat{\boldsymbol{\beta}}_{n}^{(0)}$ is the reconstructed observed amplitudes. By multiplying the reconstructed amplitude vector by $\beta_{n}^{1}$ which consists of either zeros or ones (see Eq. \ref{eq:5}), the unobserved channels will be reconstructed with amplitudes of zero and the observed channels will be reconstructed with the exponential function.
% since $\beta_{n}^{1}$ contains ones for observed channels and zeros for unobserved channels.
For our VAE, instead of estimating the distribution of $a_{n}$, we directly optimize $a_{n}$ when maximizing the lower bound. We set the initial value of $a_{n}$ to the mean of the prior. Thus, $a_{n}$ can be read as a parameter of the decoder.

% One change we make to the model when implementing it as a VAE is that we directly optimize $a_{i,k}$ when maximizing the lower bound. So 
Given our inference network and decoder, the ELBO we maximize for each spike, $s_{n}$, is given by,
% x_{s_{i,k}}, y_{s_{i,k}}, z_{s_{i,k}}
\begin{align}
\log{p(\boldsymbol{\beta}_{n};a_{n})} \geq  
-\KL{q_{\Phi}(x_n,y_n,z_n)}{p_{x_n} p_{y_n} p_{z_n}} + \mathbb{E}_{q_{r\Phi}} \left[ \sum_{l=1}^L \mathcal{N}(\beta_{n,l}^{0} \vert a_{n} \exp(b \hat{\boldsymbol{r}}_{n}), I) \beta_{n,l}^{1} \right] \nonumber
\end{align}
where KL is the KL-divergence. The location priors, $p_{x_n}, p_{y_n}, p_{z_n}$, are normally distributed as described in \ref{sec:model}, with means of zero (the position of the maximum amplitude channel in the observed data) and variances of 80 (an arbitrarily high value). For more information about the architecture and training, see Appendix F.
% to flatten the distribution

\subsubsection{Stabilized Location Estimation}\label{sec:ampjit}
In this model, the channel on which the input is centered can bias the estimate of the spike location, in particular when amplitudes are small. To reduce this bias, we can create \textit{multiple inputs} for the same spike where \textit{each input is centered on a different channel}. During inference, we can average the inferred locations for each of these inputs, thus lowering the central channel bias. To this end, we introduce a hyperparameter, \textit{amplitude jitter}, where for each spike, $s_{n}$, we create multiple inputs centered on channels with peak amplitudes within a small voltage of the maximum amplitude, $\alpha_{n,m}$. We use two values for the amplitude jitter in our experiments: $0\mu V$ and $10\mu V$. When amplitude jitter is set to $0\mu V$, no averaging is performed; when amplitude jitter is set to $10\mu V$, all channels that have peak amplitudes within $10\mu V$ of $\alpha_{n,m}$ are used as inputs to the VAE and averaged during inference.

% We set the network to be 2 layers deep with ReLu nonlinearities. The hidden unit sizes in the inference network are set to be [500, 250]. We include batchnorm layers throughout the encoder to improve training and generalization. 
%  The inference network is trained normally, but during inference, the mean location posterior estimates of all inputs that belong to the same spike are averaged.

% We train the VAE with three different learning rates, $\{.0003, .001, .003\}$ and choose the learning rate that has the highest performance, although this parameter did not have a large effect on the results.

%Add encoder and decoder.
%Define the inference network (Akash LDA paper)
%Decoder

%% file: includes/7_experiments.tex
\begin{table*}
    \centering
    \setlength\tabcolsep{3pt}
    \vspace*{3px}
    \begin{tabular}{c|c|ccc}
    \hline
    \textbf{Method} & \textbf{Observed Channels} & \multicolumn{3}{c}{ \textbf{2D Avg. Spike Distance from Soma ($\mu m$)} } \\
    % \hline
    & & 10 $\mu$V & 20 $\mu$V & 30 $\mu$V \\
    \hline
    COM & 4 & 15.84$\pm$10.08 & 16.46$\pm$10.39 & 17.18$\pm$10.97 \\
    % \hline
    COM & 9 & 18.05$\pm$11.42 & 18.59$\pm$11.67 & 19.22$\pm$12.1 \\
    % \hline
    COM & 16 & 20.94$\pm$13.09 & 21.54$\pm$13.46 & 22.17$\pm$13.94 \\
    % \hline
    COM & 25 & 23.44$\pm$14.81 & 24.31$\pm$15.43 & 25.18$\pm$15.98 \\
    % \hline
    MCMC & 9-25 & 9.87$\pm$8.64 & 11.30$\pm$9.85 & 13.31$\pm$11.67 \\
    % \hline
    VAE - 0$\mu$V & 4-9 & 9.21$\pm$8.00 & 10.40$\pm$8.97 & 12.05$\pm$10.35 \\
    % \hline
    VAE - 10$\mu$V & 4-9 &\textbf{ 8.79$\pm$7.49} & \textbf{9.79$\pm$8.31} & \textbf{11.18$\pm$9.56} \\
    % \hline
    VAE - 0$\mu$V & 9-25 & 8.94$\pm$7.91 & 10.48$\pm$9.334 & 12.43$\pm$11.223 \\
    % \hline
    VAE - 10$\mu$V & 9-25 & 9.12$\pm$7.83 & 10.41$\pm$9.07 &
    12.27$\pm$10.78 \\
    \hline
    \end{tabular}
    \caption{\textit{Results for the 2D location estimates}. These results are for three simulated, square MEA datasets with noise levels ranging from 10$\mu$V-30$\mu$V. For the VAE methods in the first column, the amount of amplitude jitter used is displayed to the right (amplitude jitter is described in \ref{sec:ampjit}).}
    \label{tab:results}
\end{table*}

\section{Experiments} \label{sec:experiments}

\subsection{Datasets}
We simulate biophysically realistic ground-truth extracellular recordings to test our model against a variety of real-life complexities. The simulations are generated using the \texttt{MEArec}~\cite{buccino2019mearec} package which includes 13 layer 5 juvenile rat somatosensory cortex neuron models from the neocortical microcircuit collaboration portal \cite{ramaswamy2015neocortical}. We simulate three recordings with increasing noise levels (ranging from $10\mu V$ to $30\mu V$) for two probe geometries, a 10x10 channel square MEA with a 15 $\mu$m inter-channel distance and 64 channels from a Neuropixels probe ($\mytilde$25-40 $\mu$m inter-channel distance). Our simulations contain 40 excitatory cells and 10 inhibitory cells with random morphological subtypes, randomly distributed and rotated in 3D space around the probe (with a 20 $\mu$m minimum distance between somas). Each dataset has about 20,000 spikes in total (60 second duration). For more details on the simulation and noise model, see Appendix \ref{app:simdata}.

For the real datasets, we use public data from a Neuropixels probe~\cite{lopez201622} and from a mouse retina recorded with the BioCam4096 platform \cite{jouty2018non}. The two datasets have 6 million and 2.2 million spikes, respectively. Spike detection and sorting (with our location estimates) are done using the HerdingSpikes2 software \cite{hilgen2017unsupervised}. 

\subsection{Evaluation} \label{sec:evaluation}
Before evaluating the localization methods, we must detect the spikes from each neuron in the simulated recordings. To avoid biasing our results by our choice of detection algorithm, we assume perfect detection, extracting waveforms from channels near each spiking neuron. Once the waveforms are extracted from the recordings, we perform the data augmentation. For the square MEA we use $W = 20, 40$, which gives $L = \var{4-9}, \var{9-25}$ real channels in the observed data, respectively. For the simulated Neuropixels, we use $W = 35, 45$, which gives $L = \var{3-6}, \var{8-14}$ real channels in the observed data, respectively. Once we have the augmented dataset, we generate location estimates for all the datasets using each localization method. For straightforward comparison with center of mass, we only evaluate the 2D location estimates (in the plane of the recording device).

In the first evaluation, we assess the accuracy of each method by computing the Euclidean distance between the estimated spike locations and the associated firing neurons. We report the mean and standard deviation of the localization error for all spikes in each recording. 

In the second evaluation, we cluster the location estimates of each method using Gaussian mixture models (GMMs). The GMMs are fit with spherical covariances ranging from 45 to 75 mixture components (with a step size of 5). We report the true positive rate and accuracy for each number of mixture components when matched back to ground truth. To be clear, our use of GMMs is not a proposed spike sorting method for real data (the number of clusters is never known apriori), but rather a systematic way to evaluate whether our location estimates are more discriminable  features than those of center of mass.

In the third evaluation,  we again use GMMs to cluster the location estimates, however, this time combined with two principal components from each spike. We report the true positive rate and accuracy for each number of mixture components as before. Combining location estimates and principal components explicitly, to create a new, low-dimensional feature set, is introduced in Hilgen (2017). In this work, the principal components are whitened and then scaled with a hyperparameter, $\alpha$. To remove any bias from choosing an $\alpha$ value in our evaluation, we conduct a grid search over $\alpha = \{4, 6, 8, 10\}$ and report the best metric scores for each method.

In the fourth evaluation, we assess the generalization performance of the method by training a VAE on an extracellular dataset and then trying to infer the spike locations in another dataset where the neuron locations are different, but all other aspects are kept the same (10$\mu$V noise level, square MEA). The localization and sorting performance is then compared to that of a VAE trained directly on the second dataset and to center of mass.

Taken together, the first evaluation demonstrates how useful each method is purely as a localization tool, the second evaluation demonstrates how useful the location estimates are for spike sorting immediately after localizing, the third evaluation demonstrates how much the performance can improve given extra waveform information, and the fourth evaluation demonstrates how our method can be used across similar datasets without retraining. For all of our sorting analysis, we use SpikeInterface version 0.9.1 \cite{buccino2019spikeinterface}.

%% file: includes/8_results.tex
\begin{figure}
\centering
\includegraphics[width=1\textwidth]{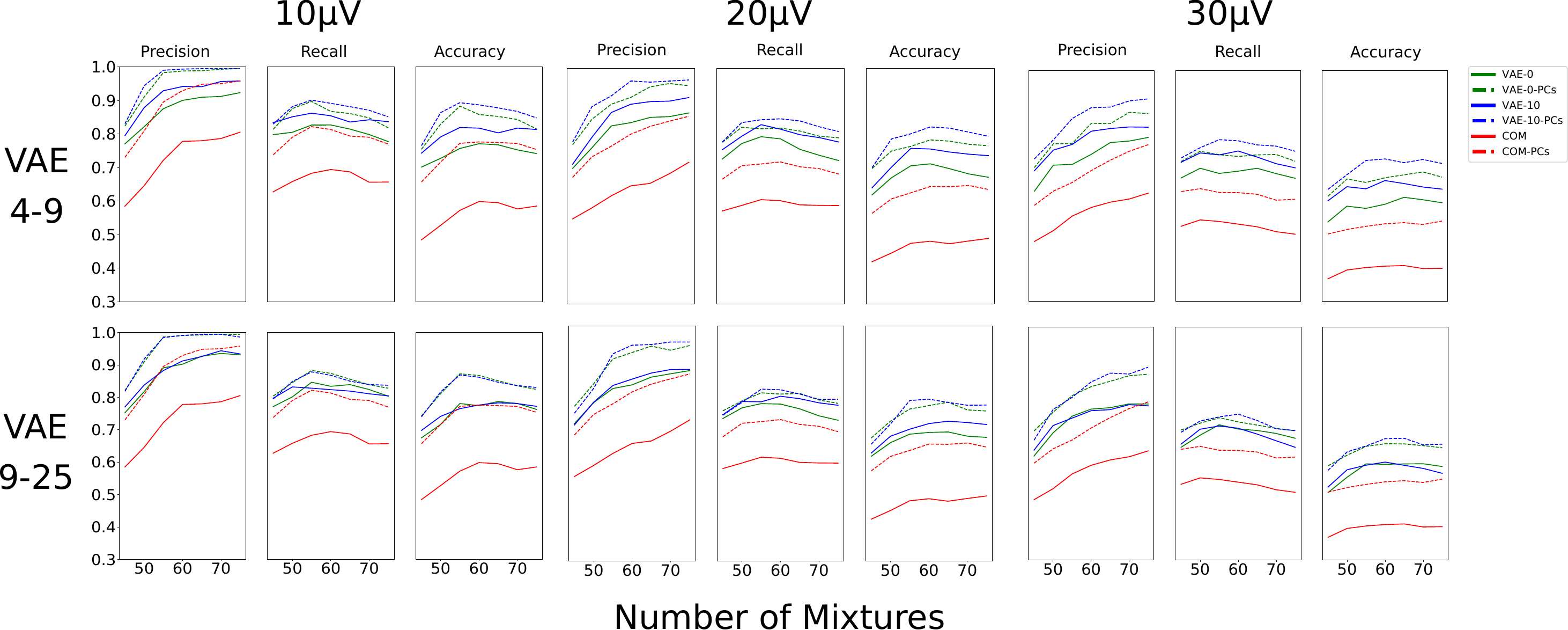}
\caption{\textit{Spike Sorting Performance on square MEA.} We compare the sorting performance of the VAE localization method and the COM localization method with and without principal components across all noise levels. For the VAE, we include the results with 0$\mu$V and 10$\mu$V amplitude jitter and with different amounts of observed channels (4-9 and 9-25). For COM, we plot the highest sorting performance (25 observed channels). The test data set has 50 neurons.}
\label{fig:sortingperformance}
\end{figure}

\subsection{Results}

Table~\ref{tab:results} reports the localization accuracy of the different localization methods for the square MEA with three different noise levels. Our model-based methods far outperform center of mass with any number of observed channels. As expected, introducing amplitude jitter helps lower the mean and standard deviation of the location spike distance. Using a small width of 20$\mu m$ when constructing the augmented data (4-9 observed channels) has the highest performance for the square MEA.

The location estimates for the square MEA are visualized in Figure~\ref{fig:simplot}. Recording channels are plotted as grey squares and the true soma locations are plotted as black stars. The estimated individual spike locations are colored according to their associated firing neuron identity. As can be seen in the plot, center of mass suffers both from artificial splitting of location estimates and poor performance on neurons outside the array, two areas in which the model-based approaches excel. The MCMC and VAE methods have very similar location estimates, highlighting the success of our variational inference in approximating the true posterior. See Appendix~\ref{app:neuroresults} for a location estimate plot when the VAE is trained and tested on simulated Neuropixels recordings.

In Figure~\ref{fig:sortingperformance}, spike sorting performance on the square MEA is visualized for all localization methods (with and without waveform information). Here, we only show the sorting results for center of mass on 25 observed channels, where it performs at its best. Overall, the model-based approaches have significantly higher precision, recall, and accuracy than center of mass across all noise levels and all different numbers of mixtures. This illustrates how model-based location estimates provide a much more discriminatory feature set than the location estimates from the center of mass approaches. We also find that the addition of waveform information (in the form of principal components) improves spike sorting performance for all localization methods. See Appendix~\ref{app:neuroresults} for a spike sorting performance plot when the VAE is trained and tested on simulated Neuropixels recordings.
% Interestingly, the VAE localization estimates have a higher sorting performance than the MCMC location estimates. We attribute this performance increase to the local coupling of parameters in the inference network that have a smoothing effect on the location estimates.

As shown in Appendix~\ref{app:genperform}, when our method is trained on one simulated recording, it can generalize well to another simulated recording with different neuron locations. The localization accuracy and sorting performance are only slightly lower than the VAE that is trained directly on the new recording. Our method also still outperforms center of mass on the new dataset even without training on it.

\begin{figure}
\centering
\includegraphics[width=0.85\textwidth]{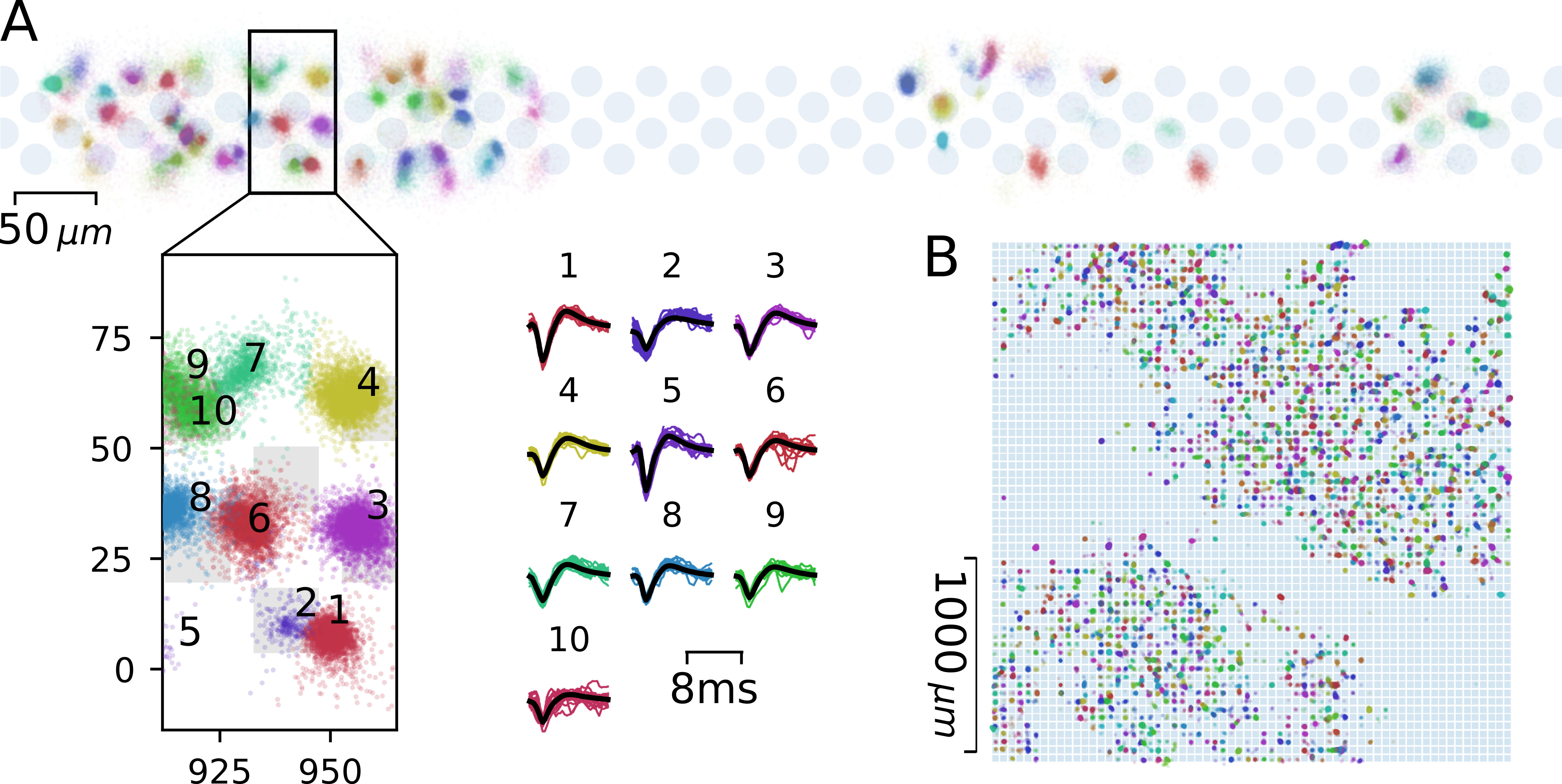}
\caption{\textit{Estimated spike locations for two real recordings.} A, Analysis of a one hour recording from an awake, head-fixed mouse with a Neuropixels probe. Spikes were detected using the HS2 package \cite{hilgen2017unsupervised}, their locations estimated using the VAE model, and clustered with mean shift, together with the first two principal components obtained from the waveforms. Shown are a large section of the probe, a magnification and corresponding spike waveforms from the clustered units. B, The same analysis performed on a recording from a mouse retina with a BioCam array from ref \cite{jouty2018non}. }
\label{fig:realplot}
% \vspace{-0.15in}
\end{figure}

Figure~\ref{fig:realplot} shows our localization method as applied to two real, large-scale extracellular datasets. In these plots, we color the location estimates based on their unit identity after spike sorting with HerdingSpikes2. These extracellular recordings do not have ground truth information as current, ground-truth recordings are limited to a few labeled neurons \cite{zanoci2019ensemble, Harris2000, henze2000intracellular, neto2016validating, yger2018spike}. Therefore, to demonstrate that the units we find likely correspond to individual neurons, we visualize waveforms from a local grouping of sorted units on the Neuropixels probe. This analysis illustrates that are method can already be applied to large-scale, real extracellular recordings.

In Appendix~\ref{app:inferencetime}, we demonstrate that the inference time for the VAE is much faster than that of MCMC, highlighting the excellent scalability of our method. The inference speed of the VAE allows for localization of one million spikes in approximately 37 seconds on a TITAN X GPU, enabling real-time analysis of large-scale extracellular datasets.

%% file: includes/9_discussion.tex
\section{Discussion}\label{sec:discussion}

Here, we introduce a Bayesian approach to spike localization using amortized variational inference. Our method significantly improves localization accuracy and spike sorting performance over the preexisting baseline while remaining scalable to the large volumes of data generated by MEAs. Scalability is particularly relevant for recordings from thousands of channels, where a single experiment may yield in the order of 100 million spikes.

% evaluated over multiple probe geometries and noise levels. 

We validate the accuracy of our model assumptions and inference scheme using biophysically realistic ground truth simulated recordings that capture much of the variability seen in real recordings. Despite the realism of our simulated recordings, there are some factors that we did not account for, including: bursting cells with event amplitude fluctuations, electrode drift, and realistic intrinsic variability of recorded spike waveforms. As these factors are difficult to model, future analysis of real recordings or advances in modeling software will help to understand possible limitations of the method.

Along with limitations of the simulated data, there are also limitations of our model. Although we assume a monopole current-source, every part of the neuronal membrane can produce action potentials \cite{buzsaki2004large}. This means that a more complicated model, such as a dipole current \cite{somogyvari2012localization}, line current-source \cite{somogyvari2012localization}, or modified ball-and-stick  \cite{ruz2014localising}, might be a better fit to the data. Since these models have only ever been used \textit{after} spike sorting, however, the extent at which they can improve localization performance \textit{before} spike sorting is unclear and is something we would like to explore in future work. Also, our model utilizes a Gaussian observation model for the spike amplitudes. In real recordings, the true noise distribution is often non-Gaussian and is better approximated by pink noise models ($\frac{1}{f}$ noise) \cite{yang2009noise}. We plan to explore more realistic observation models in future works.
% \CH{use uncertainty and differentiability?}
% \CH{post-sorting comparison}

Since our method is Bayesian, we hope to better utilize the uncertainty of the location estimates in future works. Also, as our inference network is fully differentiable, we imagine that our method can be used as a submodule in a more complex, end-to-end method. Other work indicates there is scope for constructing more complicated models to perform event detection and classification \mbox{\cite{lee2017yass}}, and to distinguish between different morphological neuron types based on their activity footprint on the array \mbox{\cite{buccino2018combining}}. Our work is thus a first step towards using amortized variational inference methods for the unsupervised analysis of complex electrophysiological recordings.

%% file: includes/appendix.tex
\section{Neuropixels Results} \label{app:neuroresults}

\begin{table*}[h]
    \centering
    \setlength\tabcolsep{3pt}
    \vspace*{3px}
    \begin{tabular}{c|c|ccc}
    \hline
    \textbf{Method} & \textbf{Observed Channels} & \multicolumn{3}{c}{ \textbf{2D Avg. Spike Distance from Soma ($\mu m$)} } \\
    & & 10 $\mu$V & 20 $\mu$V & 30 $\mu$V \\
    & & & & \\
    \hline
    COM & 4 & 23.85$\pm$12.95 & 25.16$\pm$14.21 & 26.66$\pm$15.6 \\
    % \hline
    COM & 7 & 22.81$\pm$14.04 & 24.36$\pm$15.25 & 26.11$\pm$16.63 \\
    % \hline
    COM & 12 & 26.33$\pm$15.55 & 28.26$\pm$16.66 & 30.1$\pm$17.81 \\
    % \hline
    COM & 14 & 27.83$\pm$16.26 & 30.08$\pm$17.48 & 32.03$\pm$18.57 \\
    % \hline
    MCMC & 8-14 & 14.28$\pm$ 12.68 & 16.80$\pm$15.45 & 19.74$\pm$18.30 \\
    % \hline
    VAE - 0$\mu$V & 3-6 & 14.25$\pm$12.88 & 15.74$\pm$14.88 & 18.44$\pm$17.68 \\
    % \hline
    VAE - 10$\mu$V & 3-6 & 13.10$\pm$11.04 & \textbf{15.20$\pm$13.66} & \textbf{17.68$\pm$16.38} \\
    % \hline
    VAE - 0$\mu$V & 8-14 & 13.31$\pm$12.46 & 15.63$\pm$15.51 & 18.49$\pm$18.89 \\
    % \hline
    VAE - 10$\mu$V & 8-14 & \textbf{12.91$\pm$11.41} & 15.38$\pm$14.35 & 18.14$\pm$17.55 \\
    \hline
    \end{tabular}
    \caption{\textit{Results for the 2D location estimates}. These results are for three simulated, Neuropixels datasets with noise levels ranging from 10$\mu$V-30$\mu$V. For the VAE methods in the first column, the amount of amplitude jitter used is displayed to the right (amplitude jitter is described in \ref{sec:AMI}).}
\end{table*}

\begin{figure}[ht]
\centering
\includegraphics[width=.8\textwidth]{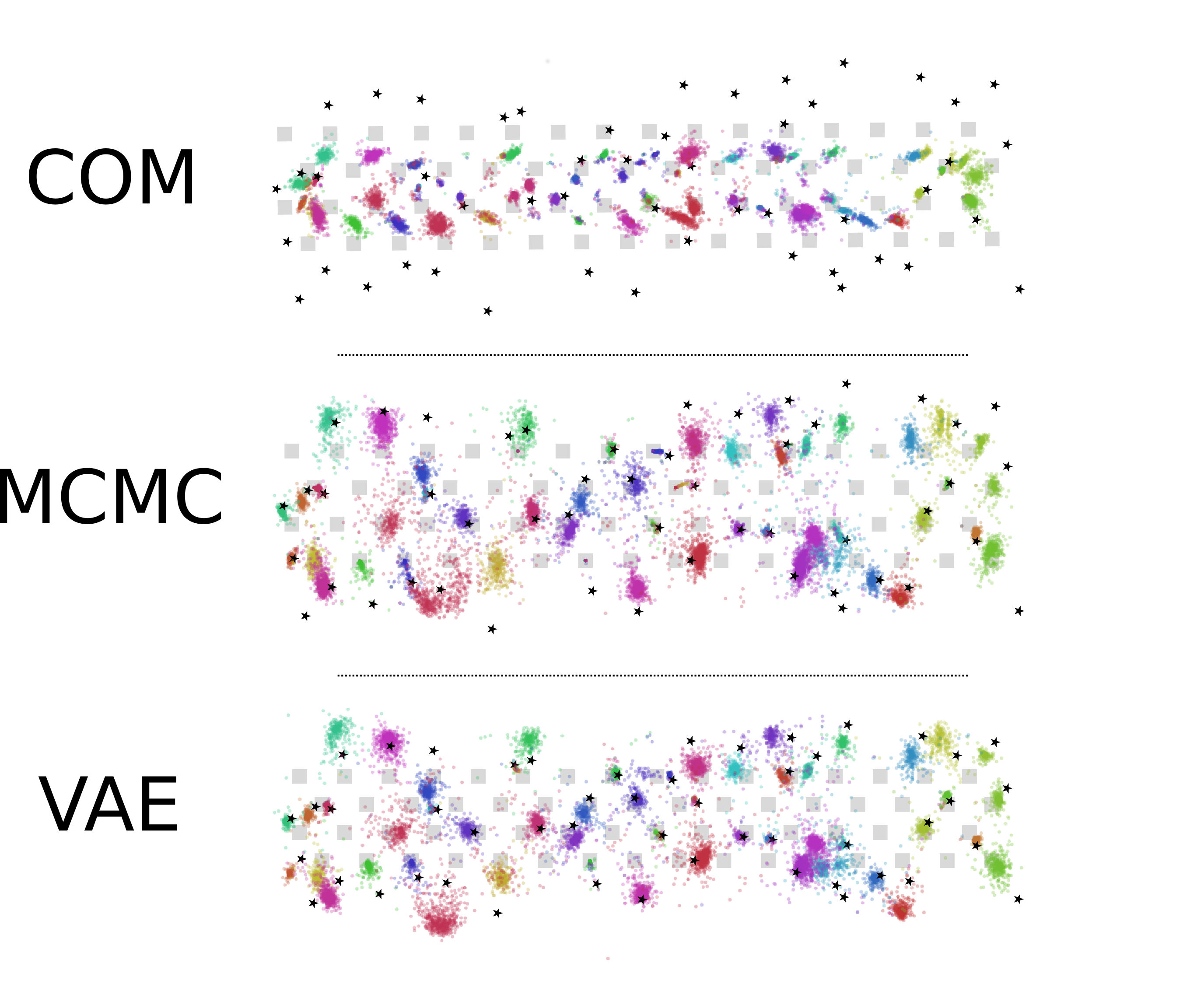}
\caption{\textit{Estimated spike locations for the different methods on a 10$\mu$V recording.} Center of mass estimates (top) are calculated using 7 channels. The MCMC estimated locations (middle) used 8-14 channels of observed amplitudes for inference, and the VAE model (bottom) was trained on 8-14 channels surrounding each spike and 0 amplitude jitter (see \ref{sec:AMI} for amplitude jitter explanation).}
\end{figure}

\begin{figure}[h]
\centering
\includegraphics[width=1\textwidth]{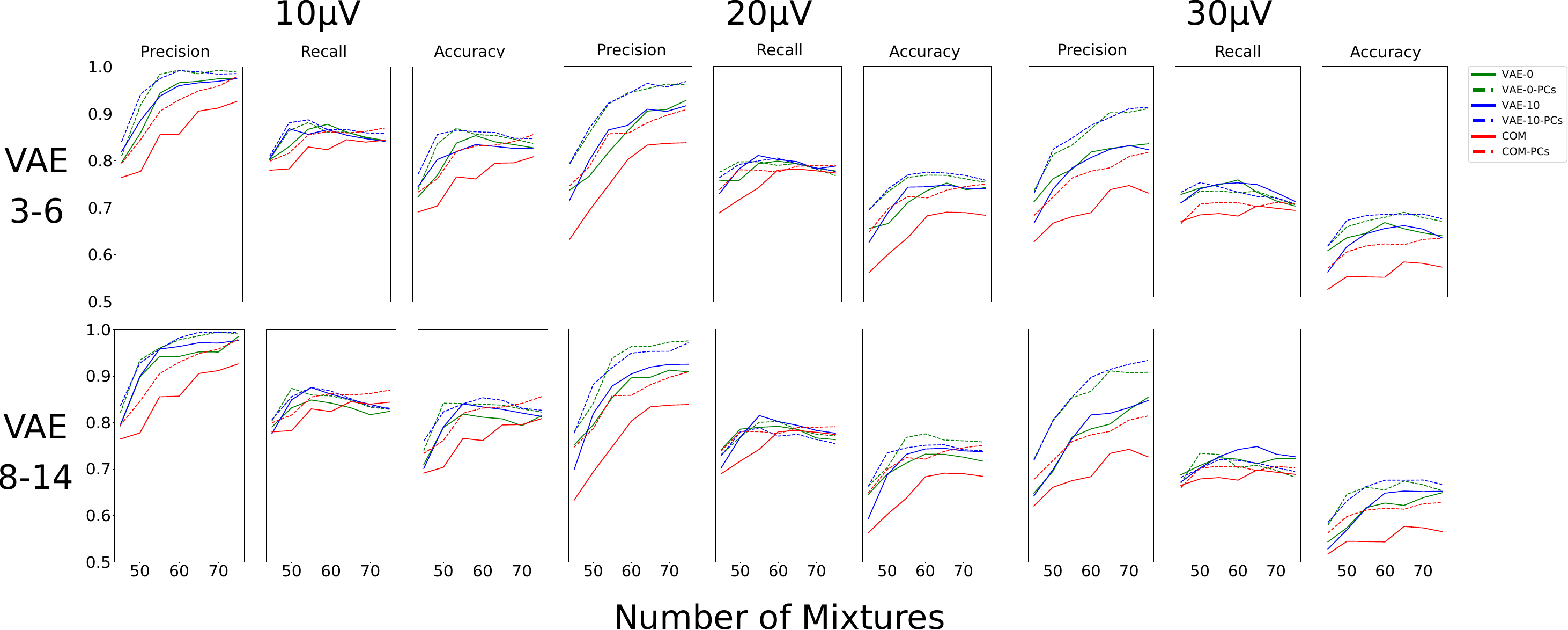}

\caption{\textit{Spike Sorting Performance on Neuropixels.} We compare the sorting performance of all localization methods with and without principal components across all noise levels. For the VAE, we include the results with and without amplitude jitter and with different amounts of real channels. For COM, we plot the highest sorting performance which was 4 observed channels. }
\end{figure}

\newpage

\section{Effect of Noise on VAE} \label{app:noise}

\begin{figure*}[h]
\centering
\includegraphics[width=.8\textwidth]{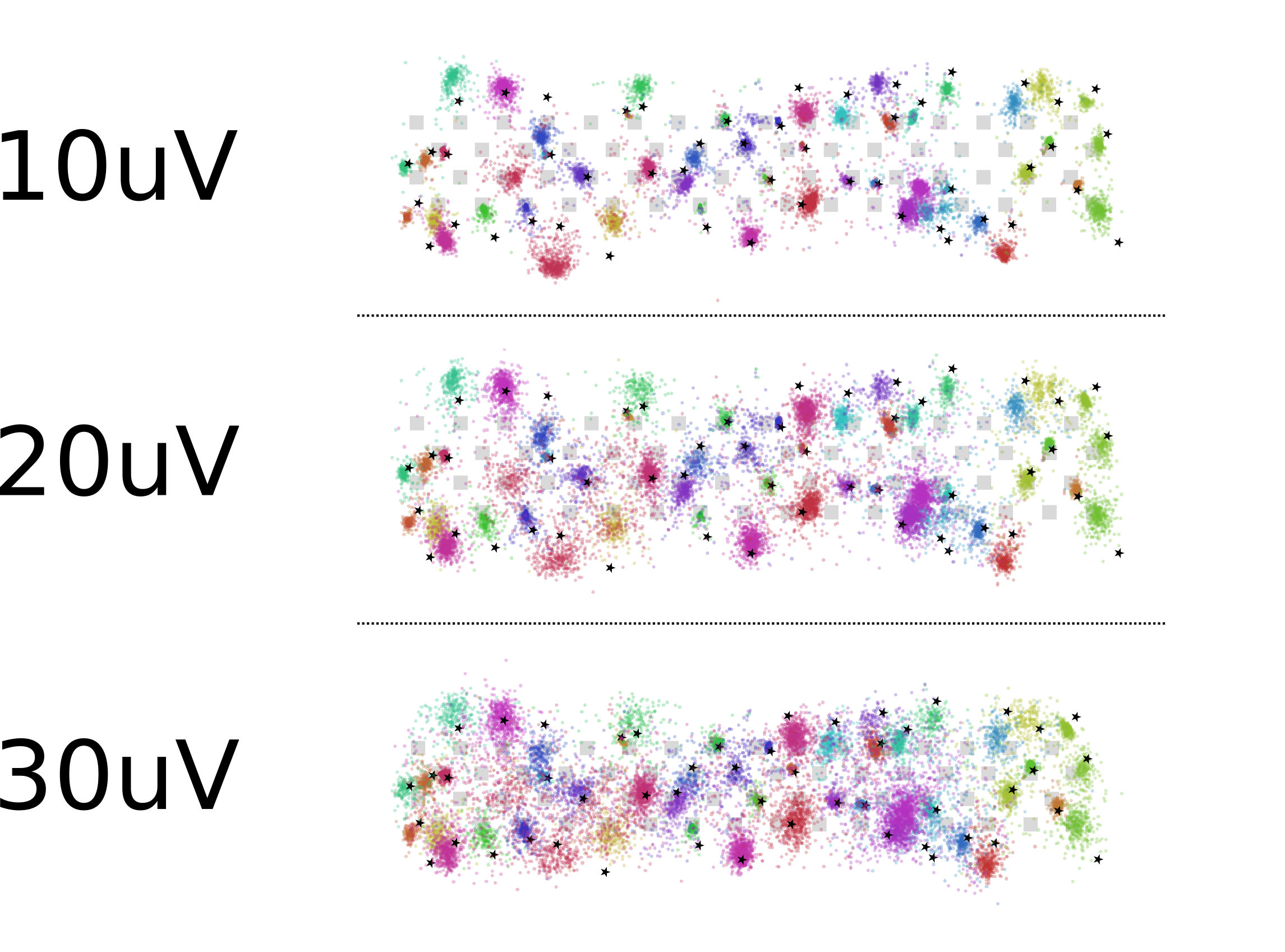}

\caption{\textit{Effect of noise on location inference for the VAE on the Neuropixels probe.} We vary the noise levels for the recording from 10$\mu$V, 20$\mu$V, and 30$\mu$V. Increasing the noise also increases the number of outliers in and the spread of the location estimates.}
\end{figure*}

\begin{figure*}[h]
\centering
\includegraphics[width=1\textwidth]{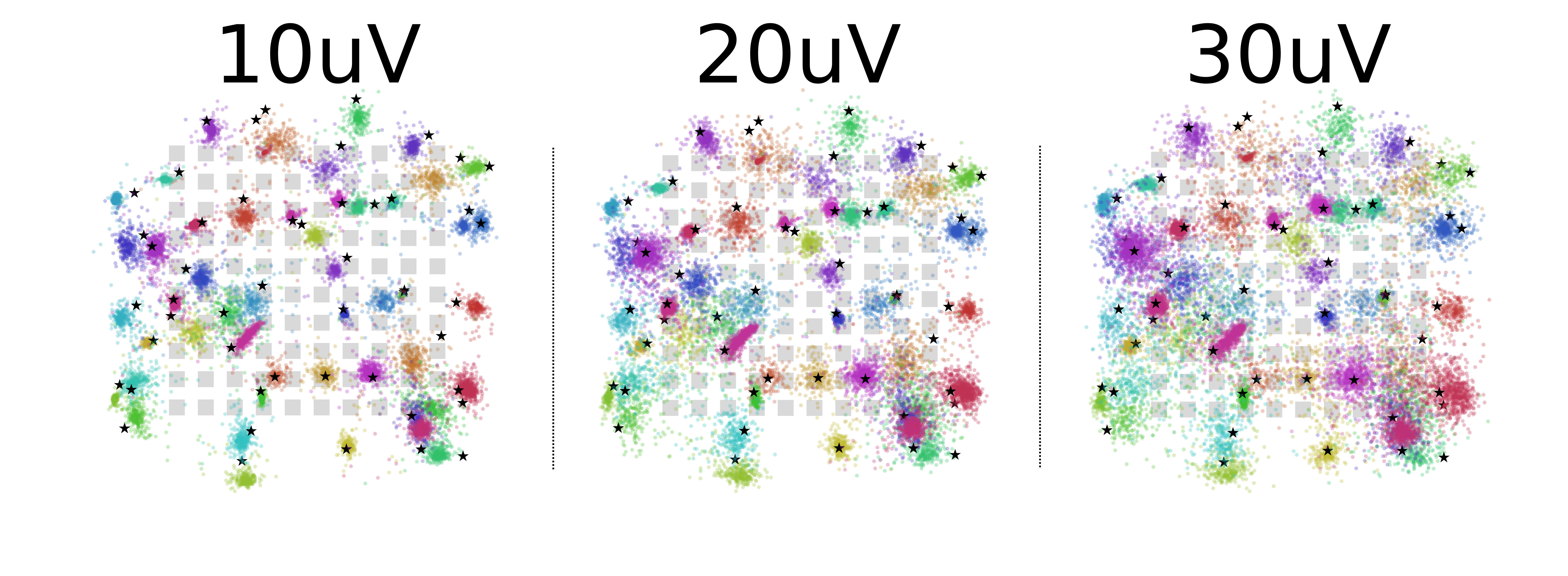}

\caption{\textit{Effect of noise on location inference for the VAE on the square MEA.} We vary the noise levels for the recording from 10$\mu$V, 20$\mu$V, and 30$\mu$V. Increasing the noise also increases the number of outliers in and the spread of the location estimates.}
\end{figure*}

\newpage

\section{Data Augmentation} \label{app:dataaug}

\begin{figure}[h]
\centering
\includegraphics[width=1\textwidth]{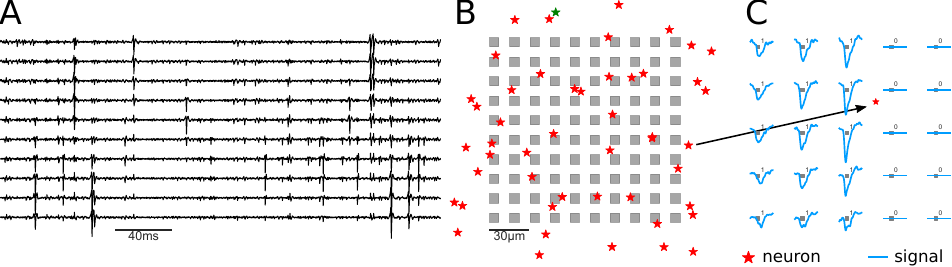}
\caption{\textit{The simulated recording set-up and example data.}
A, Example electrical traces from the MEA with recorded action potentials (spikes, negative deflections). B, The 2D layout of the simulated recording. Recording channels are indicated in grey, and the true locations of the simulated neurons in red. The traces in part A are taken from the first column of the array. Note each spike is visible in multiple channels, with a characteristic spatial decay. C, Illustration of the data augmentation procedure in cases where the spikes are detected on channels near the array boundary. A set of virtual channels is introduced, which are incapable of recording any signal, but would report non-zero amplitudes if they were present on the MEA. }
\end{figure}

\newpage

\section{Generalization Performance}\label{app:genperform}

\begin{table*}[h!]
    \centering
    \setlength\tabcolsep{3pt}
    % \label{Results Table}
    \caption{Location results for the generalization performance of a VAE trained on one 10$\mu$V, square MEA dataset and tested on another 10$\mu$V, square MEA dataset. We compare the results of this VAE to another VAE that is trained directly on the second dataset to quantify the drop in performance when generalizing between datasets. We also compare to the center of mass baselines.}
    \vspace*{3px}
    \begin{tabular}{|c|c|c|}
    \hline
    Method & Observed Channels & 2D Avg. Spike Distance from Soma (microns)	\\
    & & \\
    \hline
    COM & 4 & 16.53 $\pm$ 10.83 \\
    \hline
    COM & 9 & 18.25 $\pm$ 13.0 \\
    \hline
    COM & 16 & 20.41 $\pm$ 14.57 \\
    \hline
    COM & 25 & 22.73 $\pm$ 16.32 \\
    \hline
    VAE - 0 - Trained & 9-25 & 11.57 $\pm$ 9.88 \\
    \hline
    VAE - 0 - Inferred & 9-25 & 13.73 $\pm$ 8.01 \\
    \hline
    \end{tabular}

\end{table*}

\begin{figure*}[h]
\centering
\includegraphics[width=1\textwidth]{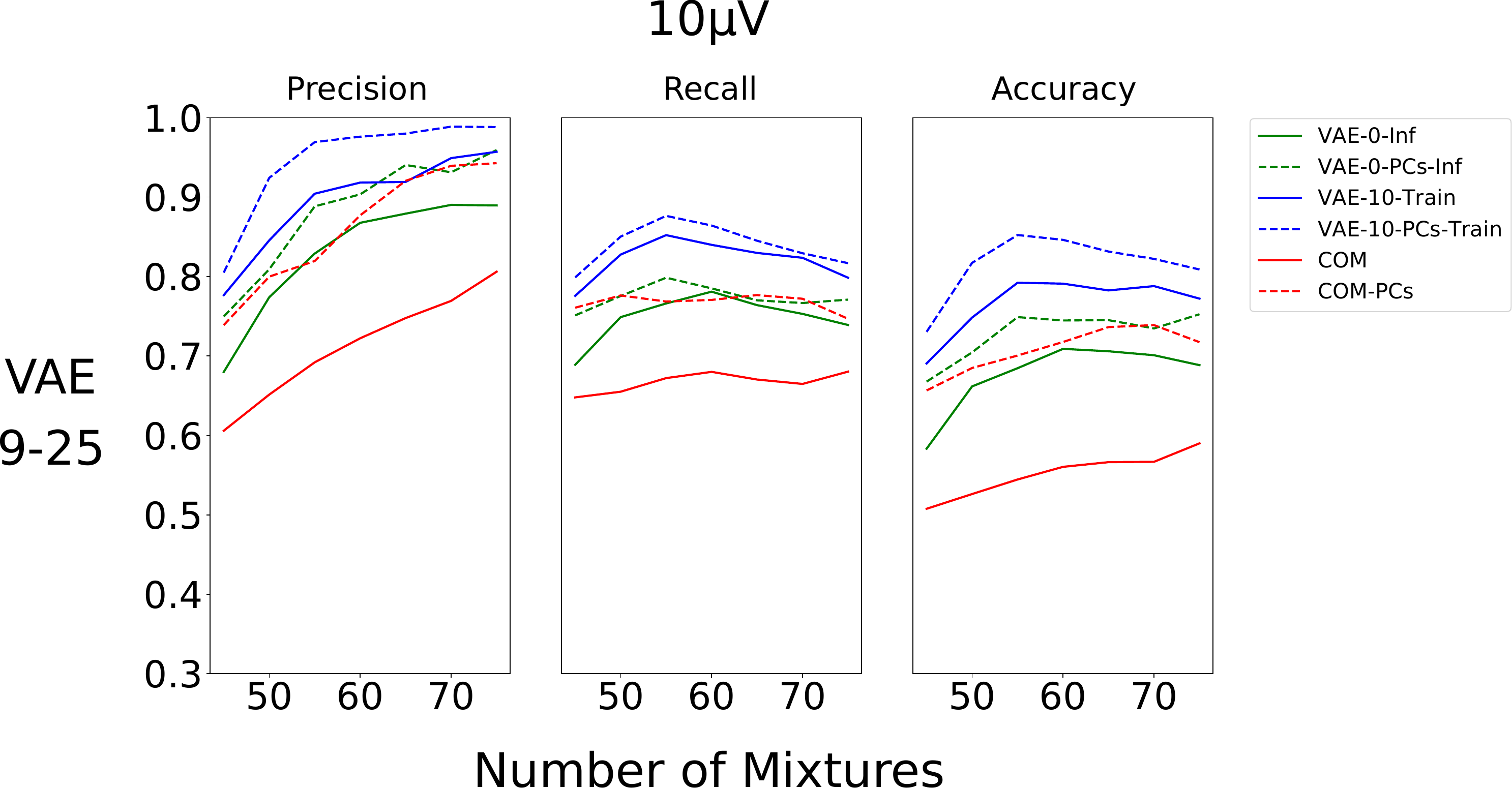}

\caption{\textit{Spike Sorting Performance Generalization.} We compare the sorting performance of the VAE localization method and the COM localization method with and without principal components across all noise levels. For the VAE, we include the results with 0$\mu$V and 10$\mu$V amplitude jitter and with different amounts of observed channels (4-9 and 9-25). For COM, we plot the highest sorting performance (25 observed channels). The test data set has 50 neurons.}
\end{figure*}

\newpage

\section{Inference Time}\label{app:inferencetime}

\begin{table*}[h!]
    \centering
    \setlength\tabcolsep{3pt}
    % \label{Results Table}
    \caption{Results for the inference time of the VAE versus HMC sampling on the dataset. We ran HMC for 10,000 iterations. The VAE was run on a TITAN X GPU.}
    \label{table:speed_results}
    \vspace*{3px}
    \begin{tabular}{|c|c|c|}
    \hline
    Method & Per Spike Inference Time (s) & Dataset Inference Time (s) \\
    \hline
    MCMC & 0.343 & 6669.0 \\
    \hline
    VAE & 0.000037 & 0.722 \\
    \hline
    \end{tabular}
\end{table*}

\section{Architecture and Training Details}

We set the inference network to be 2 layers deep with ReLU nonlinearities. The hidden unit sizes in the inference network are set to be [500, 250]. We include batchnorm layers throughout the encoder to improve training and generalization. 

We train the VAE with three different learning rates, $\{.0003, .001, .003\}$, and choose the learning rate that has the highest performance, although this parameter did not have a large effect on the results.

To ensure convergence for the simulated data, we train the network for 400 epochs on the entire dataset. For the real datasets, we train the network on a subset of the detected spikes ($\sim$100,000 spikes) and then we infer the rest of the locations.

\section{Simulated Data} \label{app:simdata}

To generate the extracellular recordings, we simulate the multi-compartment neuron models using NEURON \cite{hines1997neuron} and use the transmembrane currents to compute extracellular action potentials (EAP) with LFPy \cite{hagen2018multimodal}. EAPs are then combined with randomly generated spike trains to generate recordings. Finally, noise is added and the entire recording is filtered using a 3rd order Butterworth filter (0.3, 6 kHz). 

For the noise model, we simulate templates for 300 neurons that are far away from the recording area. These small action potentials make up the background noise of the recording and have noise levels ranging from $10\mu V$ to $30\mu V$ standard deviation for the simulated datasets. We choose this noise model because it best captures the frequency and challenges of background noise in real extracellular recordings.

For each of the three recordings on one probe geometry, we fix the neuron locations to assess the effect of noise on the location estimates for each neuron.

\section{MCMC Turing Code} \label{app:turing}
Below is the probabilistic program and inference code for the MCMC version of our method in Turing \citep{ge2018turing}.

\lstdefinelanguage{julia}
{
  keywordsprefix=\@,
  morekeywords={
    exit,whos,edit,load,is,isa,isequal,typeof,tuple,ntuple,uid,hash,finalizer,convert,promote,
    subtype,typemin,typemax,realmin,realmax,sizeof,eps,promote_type,method_exists,applicable,
    invoke,dlopen,dlsym,system,error,throw,assert,new,Inf,Nan,pi,im,begin,while,for,in,return,
    break,continue,macro,quote,let,if,elseif,else,try,catch,end,bitstype,ccall,do,using,module,
    import,export,importall,baremodule,immutable,local,global,const,Bool,Int,Int8,Int16,Int32,
    Int64,Uint,Uint8,Uint16,Uint32,Uint64,Float32,Float64,Complex64,Complex128,Any,Nothing,None,
    function,type,typealias,abstract
  },
  sensitive=true,
  morecomment=[l]{\#},
  morestring=[b]',
  morestring=[b]" 
}

\definecolor{mygray}{RGB}{128,128,128}
\definecolor{myblue}{RGB}{0, 0, 255}
\definecolor{myolivegreen}{RGB}{186, 184, 108}
\definecolor{mymaroon}{RGB}{128, 0, 0}

\lstset{
    language=julia,
    basicstyle=\small\ttfamily, 
    columns=fullflexible, % make sure to use fixed-width font, CM typewriter is NOT fixed width
    numbers=left, 
    numberstyle=\small\ttfamily\color{mygray},
    stepnumber=1,              
    numbersep=10pt, 
    numberfirstline=true, 
    numberblanklines=true, 
    tabsize=4,
    lineskip=-1.5pt,
    extendedchars=true,
    breaklines=true,        
    keywordstyle=\color{myblue}\bfseries,
    identifierstyle=, % using emph or index keywords
    commentstyle=\sffamily\color{myolivegreen},
    stringstyle=\color{mymaroon},
    showstringspaces=false,
    showtabs=false,
    upquote=false
}

\begin{lstlisting}[language=julia]
using Turing

# Define model
@model BayesianExpSpike(x_0, y_0, z_0, p_mean, p_std, locs, amps) = begin
    S_a = 1
    n_x ~ Normal(x_0, 80)
    n_y ~ Normal(y_0, 80)
    n_z ~ Normal(z_0, 80)
    
    a ~ Normal(p_mean[1], p_std[1])
    b = 0.035
    
    r = sqrt.(sum((abs.(locs .- [n_x; n_y; n_z;])).^2; dims=1))
    amps ~ MvNormal(vec(-a .* exp.(-b .* r)), S_a^0.5)
end

# Load data
real_channel_locs = ...
real_amps = ...
min_amp = ...
p_m = [2 * abs(min_amp)]
p_s = [50]

# Feed data into model
model = BayesianExpSpike(0, 0, 0, p_m, p_s, real_channel_locs, real_amps)

# Sampling
chn = sample(model, HMC(0.01, 10), 10_000)
\end{lstlisting}

%% file: 0_main.bbl
\begin{thebibliography}{56}
\providecommand{\natexlab}[1]{#1}
\providecommand{\EM}{\em}
\providecommand{\RNtxt}{\relax}
\RNtxt{}

\bibitem[Ballini et~al.(2014)M.~Ballini, J.~Muller, P.~Livi, Y.~Chen, U.~Frey,
  A.~Stettler, A.~Shadmani, V.~Viswam, I.~L. Jones, D.~Jackel, M.~Radivojevic,
  M.~K. Lewandowska, W.~Gong, M.~Fiscella, D.~J. Bakkum, F.~Heer,
  A.~Hierlemann]{Ballini2014}
{\EM Ballini Marco, Muller Jan, Livi Paolo, Chen Yihui, Frey Urs, Stettler
  Alexander, Shadmani Amir, Viswam Vijay, Jones Ian~Lloyd, Jackel David,
  Radivojevic Milos, Lewandowska Marta~K., Gong Wei, Fiscella Michele, Bakkum
  Douglas~J., Heer Flavio, Hierlemann Andreas}.
\newblock {A 1024-channel CMOS microelectrode array with 26,400 electrodes for
  recording and stimulation of electrogenic cells in vitro}
  \allowbreak\newblock// IEEE Journal of Solid-State Circuits. 2014. 49, 11.
  2705--2719.

\bibitem[Berdondini et~al.(2005)L.~Berdondini, P.~D. van~der Wal, O.~Guenat,
  N.~F. de~Rooij, M.~Koudelka-Hep, P.~Seitz, R.~Kaufmann, P.~Metzler, N.~Blanc,
  S.~Rohr]{Berdondini2005a}
{\EM Berdondini L, Wal P~D van~der, Guenat O, Rooij N~F de, Koudelka-Hep M,
  Seitz P, Kaufmann R, Metzler P, Blanc N, Rohr S}.
\newblock {High-density electrode array for imaging in vitro
  electrophysiological activity.} \allowbreak\newblock// Biosensors {\&}
  Bioelectronics. jul 2005. 21, 1. 167--74.

\bibitem[Blanche et~al.(2005)T.~J. Blanche, M.~A. Spacek, J.~F. Hetke, N.~V.
  Swindale]{blanche2005polytrodes}
{\EM Blanche Timothy~J, Spacek Martin~A, Hetke Jamille~F, Swindale Nicholas~V}.
\newblock Polytrodes: high-density silicon electrode arrays for large-scale
  multiunit recording \allowbreak\newblock// Journal of neurophysiology. 2005.
  93, 5. 2987--3000.

\bibitem[Buccino, Einevoll(2019)A.~P. Buccino, G.~T.
  Einevoll]{buccino2019mearec}
{\EM Buccino Alessio~P, Einevoll Gaute~T}.
\newblock MEArec: a fast and customizable testbench simulator for ground-truth
  extracellular spiking activity \allowbreak\newblock// bioRxiv. 2019.  691642.

\bibitem[Buccino et~al.(2019)A.~P. Buccino, C.~L. Hurwitz, J.~Magland,
  S.~Garcia, J.~H. Siegle, R.~Hurwitz, M.~H. Hennig]{buccino2019spikeinterface}
{\EM Buccino Alessio~P, Hurwitz Cole~L, Magland Jeremy, Garcia Samuel, Siegle
  Joshua~H, Hurwitz Roger, Hennig Matthias~H}.
\newblock SpikeInterface, a unified framework for spike sorting
  \allowbreak\newblock// BioRxiv. 2019.  796599.

\bibitem[Buccino et~al.(2018)A.~P. Buccino, M.~Kordovan, T.~V.~B. Ness,
  B.~Merkt, P.~D. H{\"a}fliger, M.~Fyhn, G.~Cauwenberghs, S.~Rotter, G.~T.
  Einevoll]{buccino2018combining}
{\EM Buccino Alessio~Paolo, Kordovan Michael, Ness Torbj{\o}rn V~B{\ae}k{\o},
  Merkt Benjamin, H{\"a}fliger Philipp~Dominik, Fyhn Marianne, Cauwenberghs
  Gert, Rotter Stefan, Einevoll Gaute~T}.
\newblock Combining biophysical modeling and deep learning for multi-electrode
  array neuron localization and classification \allowbreak\newblock// Journal
  of neurophysiology. 2018.

\bibitem[Buzs{\'a}ki(2004)G.~Buzs{\'a}ki]{buzsaki2004large}
{\EM Buzs{\'a}ki Gy{\"o}rgy}.
\newblock Large-scale recording of neuronal ensembles \allowbreak\newblock//
  Nature neuroscience. 2004. 7, 5. 446.

\bibitem[Carlson, Carin(2019)D.~Carlson, L.~Carin]{carlson2019continuing}
{\EM Carlson David, Carin Lawrence}.
\newblock Continuing progress of spike sorting in the era of big data
  \allowbreak\newblock// Current opinion in neurobiology. 2019. 55. 90--96.

\bibitem[Chelaru, Jog(2005)M.~I. Chelaru, M.~S. Jog]{chelaru2005spike}
{\EM Chelaru Mircea~I, Jog Mandar~S}.
\newblock Spike source localization with tetrodes \allowbreak\newblock//
  Journal of neuroscience methods. 2005. 142, 2. 305--315.

\bibitem[Chung et~al.(2017)J.~E. Chung, J.~F. Magland, A.~H. Barnett, V.~M.
  Tolosa, A.~C. Tooker, K.~Y. Lee, K.~G. Shah, S.~H. Felix, L.~M. Frank, L.~F.
  Greengard]{chung2017fully}
{\EM Chung Jason~E, Magland Jeremy~F, Barnett Alex~H, Tolosa Vanessa~M, Tooker
  Angela~C, Lee Kye~Y, Shah Kedar~G, Felix Sarah~H, Frank Loren~M, Greengard
  Leslie~F}.
\newblock A fully automated approach to spike sorting \allowbreak\newblock//
  Neuron. 2017. 95, 6. 1381--1394.

\bibitem[Dayan et~al.(1995)P.~Dayan, G.~E. Hinton, R.~M. Neal, R.~S.
  Zemel]{dayan1995helmholtz}
{\EM Dayan Peter, Hinton Geoffrey~E, Neal Radford~M, Zemel Richard~S}.
\newblock The helmholtz machine \allowbreak\newblock// Neural computation.
  1995. 7, 5. 889--904.

\bibitem[Dimitriadis et~al.(2018)G.~Dimitriadis, J.~P. Neto, A.~Aarts,
  A.~Alexandru, M.~Ballini, F.~Battaglia, L.~Calcaterra, F.~David, R.~Fiath,
  J.~Frazao, et~al.]{dimitriadis2018not}
{\EM Dimitriadis George, Neto Joana~P, Aarts Arno, Alexandru Andrei, Ballini
  Marco, Battaglia Francesco, Calcaterra Lorenza, David Francois, Fiath
  Richard, Frazao Joao, others }.
\newblock Why not record from every channel with a CMOS scanning probe?
  \allowbreak\newblock// bioRxiv. 2018.  275818.

\bibitem[Eversmann et~al.(2003)B.~Eversmann, M.~Jenkner, F.~Hofmann, C.~Paulus,
  R.~Brederlow, B.~Holzapfl, P.~Fromherz, M.~Merz, M.~Brenner, M.~Schreiter,
  R.~Gabl, K.~Plehnert, M.~Steinhauser, G.~Eckstein, D.~Schmitt-landsiedel,
  R.~Thewes]{Eversmann2003}
{\EM Eversmann Bj{\"{o}}rn, Jenkner Martin, Hofmann Franz, Paulus Christian,
  Brederlow Ralf, Holzapfl Birgit, Fromherz Peter, Merz Matthias, Brenner
  Markus, Schreiter Matthias, Gabl Reinhard, Plehnert Kurt, Steinhauser
  Michael, Eckstein Gerald, Schmitt-landsiedel Doris, Thewes Roland}.
\newblock {A 128 128 CMOS Biosensor Array for Extracellular Recording of Neural
  Activity} \allowbreak\newblock// IEEE Journal of Solid-State Circuits. 2003.
  38, 12. 2306--2317.

\bibitem[Frey et~al.(2010)U.~Frey, J.~Sedivy, F.~Heer, R.~Pedron, M.~Ballini,
  J.~Mueller, D.~Bakkum, S.~Hafizovic, F.~D. Faraci, F.~Greve, K.~U. Kirstein,
  A.~Hierlemann]{Frey2010}
{\EM Frey Urs, Sedivy Jan, Heer Flavio, Pedron Rene, Ballini Marco, Mueller
  Jan, Bakkum Douglas, Hafizovic Sadik, Faraci Francesca~D., Greve Frauke,
  Kirstein Kay~Uwe, Hierlemann Andreas}.
\newblock {Switch-matrix-based high-density microelectrode array in CMOS
  technology} \allowbreak\newblock// IEEE Journal of Solid-State Circuits.
  2010. 45, 2. 467--482.

\bibitem[Ge et~al.(2018)H.~Ge, K.~Xu, Z.~Ghahramani]{ge2018turing}
{\EM Ge~Hong, Xu~Kai, Ghahramani Zoubin}.
\newblock Turing: Composable inference for probabilistic programming
  \allowbreak\newblock// International Conference on Artificial Intelligence
  and Statistics. 2018.  1682--1690.

\bibitem[Gray et~al.(1995)C.~M. Gray, P.~E. Maldonado, M.~Wilson,
  B.~McNaughton]{gray1995tetrodes}
{\EM Gray Charles~M, Maldonado Pedro~E, Wilson Mathew, McNaughton Bruce}.
\newblock Tetrodes markedly improve the reliability and yield of multiple
  single-unit isolation from multi-unit recordings in cat striate cortex
  \allowbreak\newblock// Journal of Neuroscience Methods. 1995. 63, 1-2.
  43--54.

\bibitem[Hagen et~al.(2018)E.~Hagen, S.~N{\ae}ss, T.~V. Ness, G.~T.
  Einevoll]{hagen2018multimodal}
{\EM Hagen Espen, N{\ae}ss Solveig, Ness Torbj{\o}rn~V, Einevoll Gaute~T}.
\newblock Multimodal Modeling of Neural Network Activity: Computing LFP, ECoG,
  EEG, and MEG Signals With LFPy 2.0 \allowbreak\newblock// Frontiers in
  Neuroinformatics. 2018. 12.

\bibitem[Hagen et~al.(2015)E.~Hagen, T.~V. Ness, A.~Khosrowshahi,
  C.~S{\o}rensen, M.~Fyhn, T.~Hafting, F.~Franke, G.~T. Einevoll]{Hagen2015}
{\EM Hagen Espen, Ness Torbj{\o}rn~V., Khosrowshahi Amir, S{\o}rensen
  Christina, Fyhn Marianne, Hafting Torkel, Franke Felix, Einevoll Gaute~T.}
\newblock {ViSAPy: A Python tool for biophysics-based generation of virtual
  spiking activity for evaluation of spike-sorting algorithms}
  \allowbreak\newblock// Journal of Neuroscience Methods. 2015. 245. 182--204.

\bibitem[Harris et~al.(2000)K.~D. Harris, D.~A. Henze, J.~Csicsvari, H.~Hirase,
  G.~Buzs{\'{a}}ki]{Harris2000}
{\EM Harris Kenneth~D, Henze Darrell~A, Csicsvari J, Hirase H, Buzs{\'{a}}ki
  G}.
\newblock {Accuracy of tetrode spike separation as determined by simultaneous
  intracellular and extracellular measurements.} \allowbreak\newblock// Journal
  of Neurophysiololgy. 2000. 84, 1. 401--414.

\bibitem[Hennig et~al.(2018)M.~H. Hennig, C.~Hurwitz,
  M.~Sorbaro]{hennig2018scaling}
{\EM Hennig Matthias~H, Hurwitz Cole, Sorbaro Martino}.
\newblock Scaling Spike Detection and Sorting for Next Generation
  Electrophysiology \allowbreak\newblock// arXiv preprint arXiv:1809.01051.
  2018.

\bibitem[Henze et~al.(2000)D.~A. Henze, Z.~Borhegyi, J.~Csicsvari, A.~Mamiya,
  K.~D. Harris, G.~Buzsaki]{henze2000intracellular}
{\EM Henze Darrell~A, Borhegyi Zsolt, Csicsvari Jozsef, Mamiya Akira, Harris
  Kenneth~D, Buzsaki Gyorgy}.
\newblock Intracellular features predicted by extracellular recordings in the
  hippocampus in vivo \allowbreak\newblock// Journal of neurophysiology. 2000.
  84, 1. 390--400.

\bibitem[Hilgen et~al.(2017)G.~Hilgen, M.~Sorbaro, S.~Pirmoradian, J.-O.
  Muthmann, I.~E. Kepiro, S.~Ullo, C.~J. Ramirez, A.~P. Encinas, A.~Maccione,
  L.~Berdondini, et~al.]{hilgen2017unsupervised}
{\EM Hilgen Gerrit, Sorbaro Martino, Pirmoradian Sahar, Muthmann Jens-Oliver,
  Kepiro Ibolya~Edit, Ullo Simona, Ramirez Cesar~Juarez, Encinas Albert~Puente,
  Maccione Alessandro, Berdondini Luca, others }.
\newblock Unsupervised spike sorting for large-scale, high-density
  multielectrode arrays \allowbreak\newblock// Cell Reports. 2017. 18, 10.
  2521--2532.

\bibitem[Hines, Carnevale(1997)M.~L. Hines, N.~T. Carnevale]{hines1997neuron}
{\EM Hines Michael~L, Carnevale Nicholas~T}.
\newblock The NEURON simulation environment \allowbreak\newblock// Neural
  Computation. 1997. 9, 6. 1179--1209.

\bibitem[Jouty et~al.(2018)J.~Jouty, G.~Hilgen, E.~Sernagor, M.~H.
  Hennig]{jouty2018non}
{\EM Jouty Jonathan, Hilgen Gerrit, Sernagor Evelyne, Hennig Matthias~H}.
\newblock Non-parametric physiological classification of retinal ganglion cells
  in the mouse retina \allowbreak\newblock// Frontiers in Cellular
  Neuroscience. 2018. 12. 481.

\bibitem[Jun et~al.(2017{\natexlab{a}})J.~J. Jun, N.~A. Steinmetz, J.~H.
  Siegle, D.~J. Denman, M.~Bauza, B.~Barbarits, A.~K. Lee, C.~A. Anastassiou,
  A.~Andrei, {\c{C}}.~Ayd{\i}n, et~al.]{jun2017fully}
{\EM Jun James~J, Steinmetz Nicholas~A, Siegle Joshua~H, Denman Daniel~J, Bauza
  Marius, Barbarits Brian, Lee Albert~K, Anastassiou Costas~A, Andrei
  Alexandru, Ayd{\i}n {\c{C}}a{\u{g}}atay, others }.
\newblock Fully integrated silicon probes for high-density recording of neural
  activity \allowbreak\newblock// Nature. 2017{\natexlab{a}}. 551, 7679. 232.

\bibitem[Jun et~al.(2017{\natexlab{b}})J.~J. Jun, C.~Mitelut, C.~Lai,
  S.~Gratiy, C.~Anastassiou, T.~D. Harris]{jun2017real}
{\EM Jun James~Jaeyoon, Mitelut Catalin, Lai Chongxi, Gratiy Sergey,
  Anastassiou Costas, Harris Timothy~D}.
\newblock Real-time spike sorting platform for high-density extracellular
  probes with ground-truth validation and drift correction
  \allowbreak\newblock// bioRxiv. 2017{\natexlab{b}}.  101030.

\bibitem[Kelly et~al.(2007)R.~C. Kelly, M.~A. Smith, J.~M. Samonds, A.~Kohn,
  A.~Bonds, J.~A. Movshon, T.~S. Lee]{kelly2007comparison}
{\EM Kelly Ryan~C, Smith Matthew~A, Samonds Jason~M, Kohn Adam, Bonds AB,
  Movshon J~Anthony, Lee Tai~Sing}.
\newblock Comparison of recordings from microelectrode arrays and single
  electrodes in the visual cortex \allowbreak\newblock// Journal of
  Neuroscience. 2007. 27, 2. 261--264.

\bibitem[Kingma, Welling(2013)D.~P. Kingma, M.~Welling]{kingma2013auto}
{\EM Kingma Diederik~P, Welling Max}.
\newblock Auto-encoding variational bayes \allowbreak\newblock// arXiv preprint
  arXiv:1312.6114. 2013.

\bibitem[Kubo et~al.(2008)T.~Kubo, N.~Katayama, A.~Karashima,
  M.~Nakao]{kubo20083d}
{\EM Kubo Takashi, Katayama Norihiro, Karashima Akihiro, Nakao Mitsuyuki}.
\newblock The 3D position estimation of neurons in the hippocampus based on the
  multi-site multi-unit recordings with silicon tetrodes \allowbreak\newblock//
  2008 30th Annual International Conference of the IEEE Engineering in Medicine
  and Biology Society. 2008.  5021--5024.

\bibitem[Lee et~al.(2007)C.~W. Lee, H.~Dang, Z.~Nenadic]{lee2007efficient}
{\EM Lee Chang~Won, Dang Hieu, Nenadic Zoran}.
\newblock An efficient algorithm for current source localization with tetrodes
  \allowbreak\newblock// 2007 29th Annual International Conference of the IEEE
  Engineering in Medicine and Biology Society. 2007.  1282--1285.

\bibitem[Lee et~al.(2017)J.~H. Lee, D.~E. Carlson, H.~S. Razaghi, W.~Yao, G.~A.
  Goetz, E.~Hagen, E.~Batty, E.~Chichilnisky, G.~T. Einevoll,
  L.~Paninski]{lee2017yass}
{\EM Lee Jin~Hyung, Carlson David~E, Razaghi Hooshmand~Shokri, Yao Weichi,
  Goetz Georges~A, Hagen Espen, Batty Eleanor, Chichilnisky EJ, Einevoll
  Gaute~T, Paninski Liam}.
\newblock YASS: Yet Another Spike Sorter \allowbreak\newblock// Advances in
  Neural Information Processing Systems. 2017.  4005--4015.

\bibitem[Lopez et~al.(2016)C.~M. Lopez, S.~Mitra, J.~Putzeys, B.~Raducanu,
  M.~Ballini, A.~Andrei, S.~Severi, M.~Welkenhuysen, C.~Van~Hoof, S.~Musa,
  et~al.]{lopez201622}
{\EM Lopez Carolina~Mora, Mitra Srinjoy, Putzeys Jan, Raducanu Bogdan, Ballini
  Marco, Andrei Alexandru, Severi Simone, Welkenhuysen Marleen, Van~Hoof Chris,
  Musa Silke, others }.
\newblock 22.7 A 966-electrode neural probe with 384 configurable channels in
  0.13 $\mu$m SOI CMOS \allowbreak\newblock// Solid-State Circuits Conference
  (ISSCC), 2016 IEEE International. 2016.  392--393.

\bibitem[Mechler, Victor(2012)F.~Mechler, J.~D. Victor]{mechler2012dipole}
{\EM Mechler Ferenc, Victor Jonathan~D}.
\newblock Dipole characterization of single neurons from their extracellular
  action potentials \allowbreak\newblock// Journal of computational
  neuroscience. 2012. 32, 1. 73--100.

\bibitem[Mechler et~al.(2011)F.~Mechler, J.~D. Victor, I.~E. Ohiorhenuan, A.~M.
  Schmid, Q.~Hu]{mechler2011three}
{\EM Mechler Ferenc, Victor Jonathan~D, Ohiorhenuan Ifije~E, Schmid Anita~M,
  Hu~Qin}.
\newblock Three-dimensional localization of neurons in cortical tetrode
  recordings \allowbreak\newblock// American Journal of Physiology-Heart and
  Circulatory Physiology. 2011.

\bibitem[Miller, Wilson(2008)E.~K. Miller, M.~A. Wilson]{miller2008all}
{\EM Miller Earl~K, Wilson Matthew~A}.
\newblock All my circuits: using multiple electrodes to understand functioning
  neural networks \allowbreak\newblock// Neuron. 2008. 60, 3. 483--488.

\bibitem[M{\"{u}}ller et~al.(2015)J.~M{\"{u}}ller, M.~Ballini, P.~Livi,
  Y.~Chen, M.~Radivojevic, A.~Shadmani, V.~Viswam, I.~L. Jones, M.~Fiscella,
  R.~Diggelmann, A.~Stettler, U.~Frey, D.~J. Bakkum, A.~Hierlemann, J.~Muller,
  M.~Ballini, P.~Livi, Y.~Chen, M.~Radivojevic, A.~Shadmani, V.~Viswam, I.~L.
  Jones, M.~Fiscella, R.~Diggelmann, A.~Stettler, U.~Frey, D.~J. Bakkum,
  A.~Hierlemann]{Muller2015}
{\EM M{\"{u}}ller Jan, Ballini Marco, Livi Paolo, Chen Yihui, Radivojevic
  Milos, Shadmani Amir, Viswam Vijay, Jones Ian~L, Fiscella Michele, Diggelmann
  Roland, Stettler Alexander, Frey Urs, Bakkum Douglas~J, Hierlemann Andreas,
  Muller Jan, Ballini Marco, Livi Paolo, Chen Yihui, Radivojevic Milos,
  Shadmani Amir, Viswam Vijay, Jones Ian~L, Fiscella Michele, Diggelmann
  Roland, Stettler Alexander, Frey Urs, Bakkum Douglas~J, Hierlemann Andreas}.
\newblock {High-resolution CMOS MEA platform to study neurons at subcellular,
  cellular, and network levels} \allowbreak\newblock// Lab on a Chip. 2015. 15,
  13. 2767--2780.

\bibitem[Muthmann et~al.(2015)J.-O. Muthmann, H.~Amin, E.~Sernagor,
  A.~Maccione, D.~Panas, L.~Berdondini, U.~S. Bhalla, M.~H.
  Hennig]{Muthmann2015}
{\EM Muthmann Jens-Oliver, Amin Hayder, Sernagor Evelyne, Maccione Alessandro,
  Panas Dagmara, Berdondini Luca, Bhalla Upinder~S, Hennig Matthias~H}.
\newblock {Spike Detection for Large Neural Populations Using High Density
  Multielectrode Arrays} \allowbreak\newblock// Frontiers in Neuroinformatics.
  dec 2015. 9, December. 1--21.

\bibitem[N{\'a}dasdy et~al.(1998)Z.~N{\'a}dasdy, J.~Csicsvari, M.~Penttonen,
  J.~Hetke, K.~Wise, G.~Buzsaki]{nadasdy1998extracellular}
{\EM N{\'a}dasdy Zoltan, Csicsvari JOZSEF, Penttonen MARKKU, Hetke JAMILLE,
  Wise KENSALL, Buzsaki GY{\"O}RGY}.
\newblock Extracellular recording and analysis of neuronal activity: from
  single cells to ensembles \allowbreak\newblock// Neuronal Ensembles:
  Strategies for Recording and Decoding. 1998.  17--55.

\bibitem[Neal, others(2011)R.~M. Neal, et~al.]{neal2011mcmc}
{\EM Neal Radford~M, others }.
\newblock MCMC using Hamiltonian dynamics \allowbreak\newblock// Handbook of
  markov chain monte carlo. 2011. 2, 11. 2.

\bibitem[Neto et~al.(2016)J.~P. Neto, G.~Lopes, J.~Fraz{\~a}o, J.~Nogueira,
  P.~Lacerda, P.~Bai{\~a}o, A.~Aarts, A.~Andrei, S.~Musa, E.~Fortunato,
  et~al.]{neto2016validating}
{\EM Neto Joana~P, Lopes Gon{\c{c}}alo, Fraz{\~a}o Jo{\~a}o, Nogueira Joana,
  Lacerda Pedro, Bai{\~a}o Pedro, Aarts Arno, Andrei Alexandru, Musa Silke,
  Fortunato Elvira, others }.
\newblock Validating silicon polytrodes with paired juxtacellular recordings:
  method and dataset \allowbreak\newblock// Journal of Neurophysiology. 2016.
  116, 2. 892--903.

\bibitem[Obien et~al.(2015)M.~E.~J. Obien, K.~Deligkaris, T.~Bullmann, D.~J.
  Bakkum, U.~Frey]{Obien2015}
{\EM Obien Marie Engelene~J, Deligkaris Kosmas, Bullmann Torsten, Bakkum
  Douglas~J., Frey Urs}.
\newblock {Revealing neuronal function through microelectrode array recordings}
  \allowbreak\newblock// Frontiers in Neuroscience. 2015. 9, JAN. 423.

\bibitem[Pachitariu et~al.(2016)M.~Pachitariu, N.~A. Steinmetz, S.~N. Kadir,
  M.~Carandini, K.~D. Harris]{pachitariu2016fast}
{\EM Pachitariu Marius, Steinmetz Nicholas~A, Kadir Shabnam~N, Carandini
  Matteo, Harris Kenneth~D}.
\newblock Fast and accurate spike sorting of high-channel count probes with
  KiloSort \allowbreak\newblock// Advances in Neural Information Processing
  Systems. 2016.  4448--4456.

\bibitem[Pandarinath et~al.(2018)C.~Pandarinath, D.~J. O’Shea, J.~Collins,
  R.~Jozefowicz, S.~D. Stavisky, J.~C. Kao, E.~M. Trautmann, M.~T. Kaufman,
  S.~I. Ryu, L.~R. Hochberg, et~al.]{pandarinath2018inferring}
{\EM Pandarinath Chethan, O’Shea Daniel~J, Collins Jasmine, Jozefowicz Rafal,
  Stavisky Sergey~D, Kao Jonathan~C, Trautmann Eric~M, Kaufman Matthew~T, Ryu
  Stephen~I, Hochberg Leigh~R, others }.
\newblock Inferring single-trial neural population dynamics using sequential
  auto-encoders \allowbreak\newblock// Nature methods. 2018. ~1.

\bibitem[Prentice et~al.(2011)J.~S. Prentice, J.~Homann, K.~D. Simmons,
  G.~Tka{\v{c}}ik, V.~Balasubramanian, P.~C. Nelson]{Prentice2011}
{\EM Prentice Jason~S, Homann Jan, Simmons Kristina~D, Tka{\v{c}}ik
  Ga{\v{s}}per, Balasubramanian Vijay, Nelson Philip~C}.
\newblock {Fast, scalable, Bayesian spike identification for multi-electrode
  arrays.} \allowbreak\newblock// PloS One. jan 2011. 6, 7. e19884.

\bibitem[Ramaswamy et~al.(2015)S.~Ramaswamy, J.-D. Courcol, M.~Abdellah, S.~R.
  Adaszewski, N.~Antille, S.~Arsever, G.~Atenekeng, A.~Bilgili, Y.~Brukau,
  A.~Chalimourda, et~al.]{ramaswamy2015neocortical}
{\EM Ramaswamy Srikanth, Courcol Jean-Denis, Abdellah Marwan, Adaszewski
  Stanislaw~R, Antille Nicolas, Arsever Selim, Atenekeng Guy, Bilgili Ahmet,
  Brukau Yury, Chalimourda Athanassia, others }.
\newblock The neocortical microcircuit collaboration portal: a resource for rat
  somatosensory cortex \allowbreak\newblock// Frontiers in Neural Circuits.
  2015. 9. 44.

\bibitem[Rey et~al.(2015)H.~G. Rey, C.~Pedreira, R.~{Quian Quiroga}]{Rey2015}
{\EM Rey Hernan~Gonzalo, Pedreira Carlos, {Quian Quiroga} Rodrigo}.
\newblock {Past, present and future of spike sorting techniques}
  \allowbreak\newblock// Brain Research Bulletin. 2015. 119. 106--117.

\bibitem[Rezende et~al.(2014)D.~J. Rezende, S.~Mohamed,
  D.~Wierstra]{rezende2014stochastic}
{\EM Rezende Danilo~Jimenez, Mohamed Shakir, Wierstra Daan}.
\newblock Stochastic backpropagation and approximate inference in deep
  generative models \allowbreak\newblock// arXiv preprint arXiv:1401.4082.
  2014.

\bibitem[Ruz, Schultz(2014)I.~D. Ruz, S.~R. Schultz]{ruz2014localising}
{\EM Ruz Isabel~Delgado, Schultz Simon~R}.
\newblock Localising and classifying neurons from high density MEA recordings
  \allowbreak\newblock// Journal of neuroscience methods. 2014. 233. 115--128.
  
\bibitem[Rybkin et~al.(2021)O.~Rybkin, K.~Daniilidis,
  S.~Levine]{rybkin2021simple}
{\EM Rybkin Oleh, Daniilidis Kostas, Levine Sergey}.
\newblock Simple and effective VAE training with calibrated decoders
  \allowbreak\newblock// International Conference on Machine Learning. 2021.
  9179--9189.

\bibitem[Segev et~al.(2004)R.~Segev, J.~Goodhouse, J.~Puchalla, M.~J.
  Berry~II]{segev2004recording}
{\EM Segev Ronen, Goodhouse Joe, Puchalla Jason, Berry~II Michael~J}.
\newblock Recording spikes from a large fraction of the ganglion cells in a
  retinal patch \allowbreak\newblock// Nature Neuroscience. 2004. 7, 10. 1155.

\bibitem[Somogyv{\'a}ri et~al.(2012)Z.~Somogyv{\'a}ri, D.~Cserp{\'a}n,
  I.~Ulbert, P.~{\'E}rdi]{somogyvari2012localization}
{\EM Somogyv{\'a}ri Zolt{\'a}n, Cserp{\'a}n Dorottya, Ulbert Istv{\'a}n,
  {\'E}rdi P{\'e}ter}.
\newblock Localization of single-cell current sources based on extracellular
  potential patterns: the spike CSD method \allowbreak\newblock// European
  Journal of Neuroscience. 2012. 36, 10. 3299--3313.

\bibitem[Somogyv{\'a}ri et~al.(2005)Z.~Somogyv{\'a}ri, L.~Zal{\'a}nyi,
  I.~Ulbert, P.~{\'E}rdi]{somogyvari2005model}
{\EM Somogyv{\'a}ri Zolt{\'a}n, Zal{\'a}nyi L{\'a}szl{\'o}, Ulbert Istv{\'a}n,
  {\'E}rdi P{\'e}ter}.
\newblock Model-based source localization of extracellular action potentials
  \allowbreak\newblock// Journal of neuroscience methods. 2005. 147, 2.
  126--137.

\bibitem[Speiser et~al.(2017)A.~Speiser, J.~Yan, E.~W. Archer, L.~Buesing,
  S.~C. Turaga, J.~H. Macke]{speiser2017fast}
{\EM Speiser Artur, Yan Jinyao, Archer Evan~W, Buesing Lars, Turaga Srinivas~C,
  Macke Jakob~H}.
\newblock Fast amortized inference of neural activity from calcium imaging data
  with variational autoencoders \allowbreak\newblock// Advances in Neural
  Information Processing Systems. 2017.  4024--4034.

\bibitem[Yang et~al.(2009)Z.~Yang, Q.~Zhao, E.~Keefer, W.~Liu]{yang2009noise}
{\EM Yang Zhi, Zhao Qi, Keefer Edward, Liu Wentai}.
\newblock Noise characterization, modeling, and reduction for in vivo neural
  recording \allowbreak\newblock// Advances in neural information processing
  systems. 2009.  2160--2168.

\bibitem[Yger et~al.(2018)P.~Yger, G.~L. Spampinato, E.~Esposito, B.~Lefebvre,
  S.~Deny, C.~Gardella, M.~Stimberg, F.~Jetter, G.~Zeck, S.~Picaud,
  et~al.]{yger2018spike}
{\EM Yger Pierre, Spampinato Giulia~LB, Esposito Elric, Lefebvre Baptiste, Deny
  St{\'e}phane, Gardella Christophe, Stimberg Marcel, Jetter Florian, Zeck
  Guenther, Picaud Serge, others }.
\newblock A spike sorting toolbox for up to thousands of electrodes validated
  with ground truth recordings in vitro and in vivo \allowbreak\newblock//
  eLife. 2018. 7. e34518.

\bibitem[Yuan et~al.(2016)X.~Yuan, S.~Kim, J.~Juyon, M.~D'Urbino, T.~Bullmann,
  Y.~Chen, A.~Stettler, A.~Hierlemann, U.~Frey]{yuan2016microelectrode}
{\EM Yuan X, Kim S, Juyon J, D'Urbino M, Bullmann T, Chen Y, Stettler
  Alexander, Hierlemann Andreas, Frey Urs}.
\newblock A microelectrode array with 8,640 electrodes enabling simultaneous
  full-frame readout at 6.5 kfps and 112-channel switch-matrix readout at 20
  kS/s \allowbreak\newblock// VLSI Circuits (VLSI-Circuits), 2016 IEEE
  Symposium on. 2016.  1--2.

\bibitem[Zanoci et~al.(2019)C.~Zanoci, N.~Dehghani,
  M.~Tegmark]{zanoci2019ensemble}
{\EM Zanoci Cristian, Dehghani Nima, Tegmark Max}.
\newblock Ensemble inhibition and excitation in the human cortex: An
  Ising-model analysis with uncertainties \allowbreak\newblock// Physical
  Review E. 2019. 99, 3. 032408.

\end{thebibliography}
